\documentclass[article,onecolumn,superscriptaddress,nofootinbib,floatfix,onecolumn]{revtex4}

\usepackage{amsmath, latexsym, hyperref, graphicx, color, multirow, makecell, diagbox}
\usepackage{tabularx}
\usepackage[T1]{fontenc}

\usepackage{enumitem}
\usepackage{dcolumn}
\usepackage{bm}
\usepackage{xspace}

\usepackage{float}
\usepackage{epsfig}
\usepackage{longtable}
\usepackage{rotating}
\usepackage{ulem}
\usepackage{subfigure}
\usepackage{amssymb}
\usepackage{amsmath}
\usepackage{txfonts}
\usepackage{upgreek}
\usepackage{mathtools}
\usepackage{tabularx}
\usepackage{booktabs}

\usepackage{comment}

\usepackage{soul} 

\usepackage[capitalize]{cleveref}
\definecolor{nicered}{rgb}{.7,.1,.1}
\definecolor{nicegreen}{rgb}{.1,.5,.1}
\definecolor{darkblue}{rgb}{0,0,.5}
\hypersetup{colorlinks, citecolor=blue,linkcolor=nicered, urlcolor=darkblue}
\usepackage{multirow}
\usepackage{slashed}
\usepackage{verbatim}

\def\cO#1{{{\cal{O}}}\left(#1\right)}

\newcommand{\epem}{e^+e^-}
\newcommand{\mumu}{\mu^+\mu^-}
\newcommand{\tautau}{\tau^+\tau^-}
\newcommand{\tata}{\mathcal{T}}

\newcommand{\tataOne}{\mathcal{T}_1}

\newcommand{\gaga}{\gamma\gamma}
\providecommand{\qqbar}{q\overline{q}}
\providecommand{\qqbarprime}{q'\overline{q}'}
\providecommand{\ffbar}{f\overline{f}}
\providecommand{\ffbarprime}{f'\overline{f}'}

\providecommand{\ccbar}{c\overline{c}}
\providecommand{\lele}{\ell^{+}\,\ell^{-}}

\providecommand{\nunubar}{\nu\overline{\nu}}

\newcommand{\alphas}{\alpha_\text{ s}}

\usepackage{xspace}
\newcommand*{\eg}{e.g.,\@\xspace}
\newcommand*{\ie}{i.e.,\@\xspace}

\begin{document}

\title{Ditauonium spectroscopy}
 
\author{David~d'Enterria}\email{david.d'enterria@cern.ch}
\affiliation{CERN, EP Department, CH-1211 Geneva, Switzerland}
\author{Redamy~Perez-Ramos}\email{redamy.perez-ramos@ipsa.fr}
\affiliation{DRII-IPSA, Bis, 63 Boulevard de Brandebourg, 94200 Ivry-sur-Seine, France}
\affiliation{Laboratoire de Physique Th\'eorique et Hautes Energies (LPTHE), UMR 7589,\\ Sorbonne Universit\'e et CNRS, 4 place Jussieu, 75252 Paris Cedex 05, France}
\author{Hua-Sheng~Shao}\email{huasheng.shao@lpthe.jussieu.fr}
\affiliation{Laboratoire de Physique Th\'eorique et Hautes Energies (LPTHE), UMR 7589,\\ Sorbonne Universit\'e et CNRS, 4 place Jussieu, 75252 Paris Cedex 05, France}

\begin{abstract}
\noindent
We examine the properties of ditauonium, an exotic atom consisting of a pair of opposite-sign $\tau$ leptons bound together by the quantum electrodynamics (QED) interaction in a hydrogen-like state. The energy levels, decay modes and associated partial widths, as well as total widths and lifetimes of the ortho- and para-ditauonium states are calculated. Higher-order QED effects --including Lamb shifts, hyperfine splitting structure, and partial decay widths corrections-- are incorporated up to approximately next-to-next-to-leading-order (NNLO) accuracy. Beyond the dominant diphoton and difermion decays, the rates of rare decay channels --including Dalitz, radiative, triple-photon, double-Dalitz, four-fermion, and neutrinos final states-- are determined.
\end{abstract}

\date{\today}

\maketitle

\nopagebreak

\section{Introduction}

Opposite-charge leptons ($\ell^\pm = e^\pm, \mu^\pm, \tau^\pm$) can form transient ``onium'' bound states under their quantum electrodynamics (QED) interaction. Like for the hydrogen atom, the various states of such exotic atoms feature a very rich spectroscopic structure arising from the relative spin orientation of their two leptonic spin-$1/2$ constituents as well as from various (fine and hyperfine) relativistic and quantum corrections. 
Using the spectroscopic $n^{2S+1}L_J$ notation (for principal quantum number $n$, total spin $S=0,1$, orbital angular momentum $L=0,1,2,\dots\equiv \mathrm{S},\mathrm{P},\mathrm{D},\dots$, and total angular momentum $J=L\pm 1,L$ for $S = 1$, and $J = L$ for $S = 0$), the leptonium ground state corresponds to the lowest energy orbital 
with $J=0$~and~1 for para- and ortho-leptonium states, respectively. In the first case, spin-singlet para-leptonium states $1^1\mathrm{S}_0$ have leptonic constituents with antiparallel spins, they carry $J^{PC}=0^{-+}$ quantum numbers (where $C$ and $P$ are charge conjugation and parity, respectively) and decay preferentially into two photons. In the second case, triplet ortho-leptonium ($1^3\mathrm{S}_1$) states are composed of leptons with parallel spins, have $J^{PC}=1^{--}$, and decay into $3\gamma$ or, if kinematically accessible, into lighter $\lele$ or quark-antiquark ($\qqbar$) final states. 

Out of six possible exotic leptonic atoms, $(\epem)$, $(\mu^\pm\mathrm{e}^\mp)$, $(\mumu)$, $(\tau^\pm\mathrm{e}^\mp)$, $(\tau^\pm\mu^\mp)$, and $(\tau^+\tau^-)$, only two of them ($\epem$, positronium)~\cite{Deutsch:1951zza}, and ($\mu^\pm\mathrm{e}^\mp$, muonium)~\cite{Hughes:1960zz} have been observed to date. The most well-known leptonium system is positronium, whose spectroscopy has been thoroughly studied as a means to provide stringent tests of QED~\cite{Karshenboim:2005iy}, as well as in searches for violations of the discrete $CPT$ symmetries either singly or in various combinations~\cite{Bernreuther:1988tt,Yamazaki:2009hp}. The muonic counterpart of positronium, called dimuonium or true muonium (with the \textit{true} adjective added to avoid confusion with the muonium state), has never been observed~\cite{Brodsky:2009gx}, nor the heaviest leptonium state, true tauonium or ditauonium. This work focuses on this latter system
that has been barely investigated~\cite{Malenfant:1987tm,Malik:2008pn,Fael:2018ktm} since it was first suggested in~\cite{Moffat:1975uw,Avilez:1977ai,Avilez:1978sa}, and for which first feasibility studies for its measurement at $\epem$ and hadron colliders have been recently proposed~\cite{dEnterria:2022ysg,DdEHSS}. 

Since the tau lepton is $\sim$3500 and $\sim$17 times more massive than the electron and muon, respectively, and since all the leptonium basic properties (energy levels, decay widths) are proportional to $m_{\ell}$, ditauonium properties will be correspondingly scaled by factors of about 3500 and 17 compared to their lightest (positronium and dimuonium) siblings. In this light, the investigation of ditauonium properties can provide, first, new tests of QED and of $CPT$ symmetries at much higher masses or, equivalently, at much smaller distances compared to precision studies of other exotic atoms.
Secondly, ditauonium features enhanced sensitivity to any physics beyond the standard model (BSM) at a scale $\Lambda_\text{BSM}$ that is either suppressed by powers of $\cO{m_{\ell}/\Lambda_\text{BSM}}$ or affected by hadronic uncertainties, as is the case for, \eg\ positronium or muonic-hydrogen states, respectively. The comparison of positronium, dimuonium, and ditauonium decays can thereby provide information complementary to any potential BSM effect (such as \eg\ lepton-flavor universality violation) observed with the corresponding ``open'' leptons.

This paper presents, for the first time to our knowledge, a detailed study of the main properties of ditauonium states (energy levels, partial widths for all relevant decay modes, as well as total widths and associated lifetimes) including higher-order QED contributions. In Sections~\ref{sec:energylevels} and~\ref{sec:decays} we present, respectively, the basic leading-order (LO) expressions for all quantities, followed by their more complete and accurate results including higher-order QED corrections.
The wavefunctions of the bound states at the origin are the physical quantities of interest to perturbatively compute QED onium spectroscopy properties. The LO wavefunctions are determined by solving the nonrelativistic Schr\"odinger equation with the Coulomb potential. At LO, the square of the $n$S wavefunctions at the radial origin ($r=0$) depend on the lepton mass and QED coupling, as\footnote{Natural units, $\hslash=c=1$, are used throughout the paper.}
\begin{eqnarray}
\left|\varphi_{n\mathrm{S}}(r=0)\right|^2&=&\frac{\left(\alpha m_\tau\right)^3}{8\pi n^3}.
\end{eqnarray}
As we will see below, the zeroth-order energy levels and decay rates are proportional to $\alpha^2 m_\tau$ and $\alpha^5 m_\tau$, respectively. Virtual next-to-leading-order (NLO) and next-to-next-to-leading-order (NNLO) corrections given by one-loop Feynman diagrams corrections to the Coulomb photon, \ie\ the nonrelativistic Uehling's potential, have been theoretically calculated for the light leptonium systems. Because of its larger mass, the ditauonium system has accessible many more decay channels that are kinematically forbidden in the positronium and dimuonium cases, \ie\ ditauonium has multiple higher-order real corrections that are nonexistent for its lighter siblings. In this work, we compute for the first time the full spectroscopic properties of ditauonium, including real and virtual NLO plus (leading) NNLO corrections to the energy levels (Lamb shifts of order $\alpha^3 m_\tau$, and hyperfine splittings proportional to $\alpha^4 m_\tau$, respectively), as well as to the decay rates (organising them as functions of zeroth-order widths multiplied by $(\alpha/\pi)$ and $(\alpha/\pi)^2$ terms, respectively), for the para and ortho states separately. The paper is closed with a summary of the main results in Section~\ref{sec:summ}.

\section{Energy levels}
\label{sec:energylevels}

\begin{figure}[htpb!]
\centering
\includegraphics[width=0.9\textwidth]{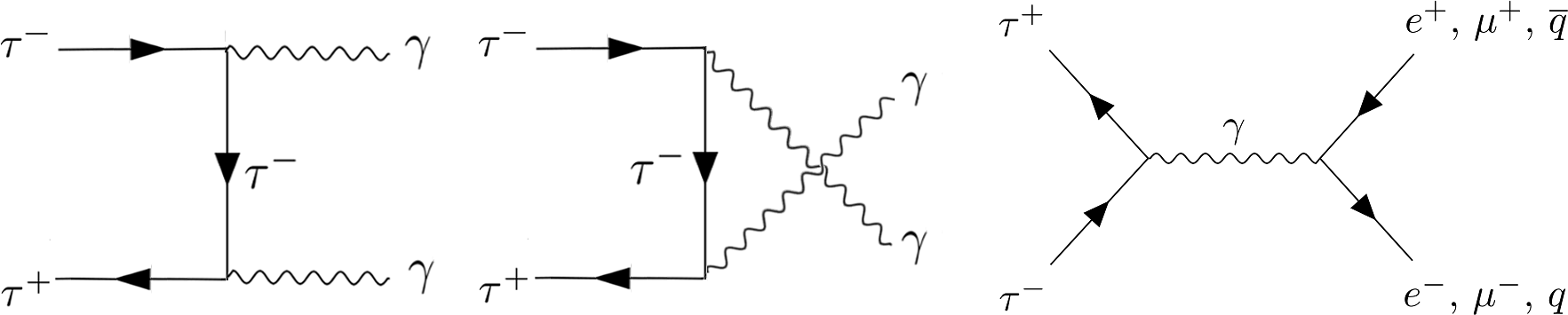}
\caption{LO diagrams of para- (left and center) and ortho- (right) ditauonium decays.
\label{fig:LO_decays_diags}}
\end{figure}

As particle-antiparticle systems, leptonia are intrinsically unstable against annihilation, and this feature makes them markedly different compared to normal atomic systems. The annihilation rates of leptonia are dependent on the overlap of the lepton and antilepton wavefunctions and, therefore, vary for different states. For leptonia in the ground state (where $L = 0$), the singlet ($S = 0$) and triplet ($S = 1$) states can only annihilate into even and odd numbers of photons respectively, because of the Landau--Yang selection rule~\cite{Landau:1948kw,Yang:1950rg} and $C$ symmetry. Thus, para-ditauonium annihilates into two real photons via a $t$-channel process (Fig.~\ref{fig:LO_decays_diags}, left and center), whereas ortho-ditauonium does it into a pair of charged fermions via an intermediate $s$-channel virtual photon (Fig.~\ref{fig:LO_decays_diags}, right), or into three real photons via the $t$-channel process (see Fig.~\ref{fig:NLO_ortho_diags} bottom right, later on).

As shown below, higher orbital states prefer to radiatively decay to the ditauonium ground state(s) rather than do it through $\tautau$ annihilation. The following two subsections present, respectively, the basic LO results and their higher-order QED and relativistic corrections, of the energy levels of the ditauonium spectrum.

\subsection{Leading-order results}

At LO, the energy levels of the exotic ditauonium atom can be described by the Schr\"odinger equation with the Coulomb potential, yielding the same Bohr expression for a hydrogen atom with reduced mass $m_{\text{red}}=m_{\tau}/2$,
\begin{equation}
E_{n} = -\frac{\alpha^2 m_\text{red}}{2n^2} = -\frac{\alpha^2 m_{\tau}}{4n^2} \approx -\frac{23.655~\mathrm{keV}}{n^2},
\label{eq:E_n}
\end{equation}
where the last numerical expression is obtained using the values of the tau lepton mass $m_\tau$ and QED coupling at zero momentum $\alpha$ listed in Table~\ref{tab:PDG}. The binding energy of the ground state ($n=1$) of true tauonium ($\tata$) is, thus, $E_{n=1} = -23.655$~keV, and its mass is 
\begin{equation}
m_{_{\tata}} = 2m_\tau + E_{n=1} = 3553.696\pm 0.240~\mbox{MeV},
\label{eq:ditauonium_mass}
\end{equation}
where the uncertainty is dominated by the current precision of the tau lepton mass~\cite{Zyla:2020zbs}. Note that the $\sim$23.65~keV binding energy of the ditauonium ground states is about ten times smaller than the current uncertainty of the central value of the ditauonium mass itself.

\begin{table}[htp!]
\centering
\caption{Values of the masses of the leptons and (approximate) constituent quarks, tau lifetime,  QED coupling, and hadronic photon vacuum polarization self-energy for $N_f=3$ quark flavours, $\Delta\alpha_\text{had}^{(3)}(m_{_{\tata}}^2)$, and $R_\text{had}(m_{_{\tata}}^2)$ ratio in $\epem$ collisions, both evaluated at the $\tata$ mass scale, used in this work~\cite{Zyla:2020zbs}. The quoted value of $\Delta\alpha_\text{had}^{(3)}(m_{_{\tata}}^2)$ is computed using \texttt{alphaQED19}~\cite{Jegerlehner:2019lxt,Proceedings:2019vxr}.
\label{tab:PDG}}
\vspace{0.1cm}
\resizebox{\textwidth}{!}{%
\begin{tabular}{cccccccccc}\hline
$m_e $ (MeV) &  $m_\mu$ (MeV) &  $m_\tau$ (MeV) & $m_u$ (MeV) &  $m_d$ (MeV) & $m_s$ (MeV) & $\tauup$~(fs) & $\alpha$ &  $\Delta\alpha_\text{had}^{(3)}(m_{_{\tata}}^2)$ & $R_\text{had}(m_{_{\tata}}^2)$ \\
$0.51099895$ &  $105.6583745$ & $1776.86 \pm 0.12$ & 335 &  340 & 490 & $290.3 \pm 0.5$ & $1/137.036$ & $\approx$0.0077 & $\approx$2.2 \\\hline
\end{tabular}
}
\end{table}

The Bohr radius of the ditauonium ground-state is $a_0 = 2/(\alpha m_{\tau})= 30.4$~fm, and its Rydberg constant amounts to $R_\infty = \alpha/(4\pi a_0) = 3.76$~keV. Namely, ditauonium is the smallest of all leptonium atoms, and has the largest ``photon ionization'' energy among them, \ie\ it is the most strongly bound of all leptonia. The velocity of each tau in the $n$-th Bohr orbit is $\beta=1/(n\,m_{\tau} a_0)=\alpha/(2n)$, which justifies the use of nonrelativistic bound-state perturbation theory (NRQED)~\cite{Caswell:1985ui} to calculate its properties as commonly done for the lighter positronium and dimuonium systems.

\begin{figure}[htpb!]
\centering
\includegraphics[width=0.75\textwidth]{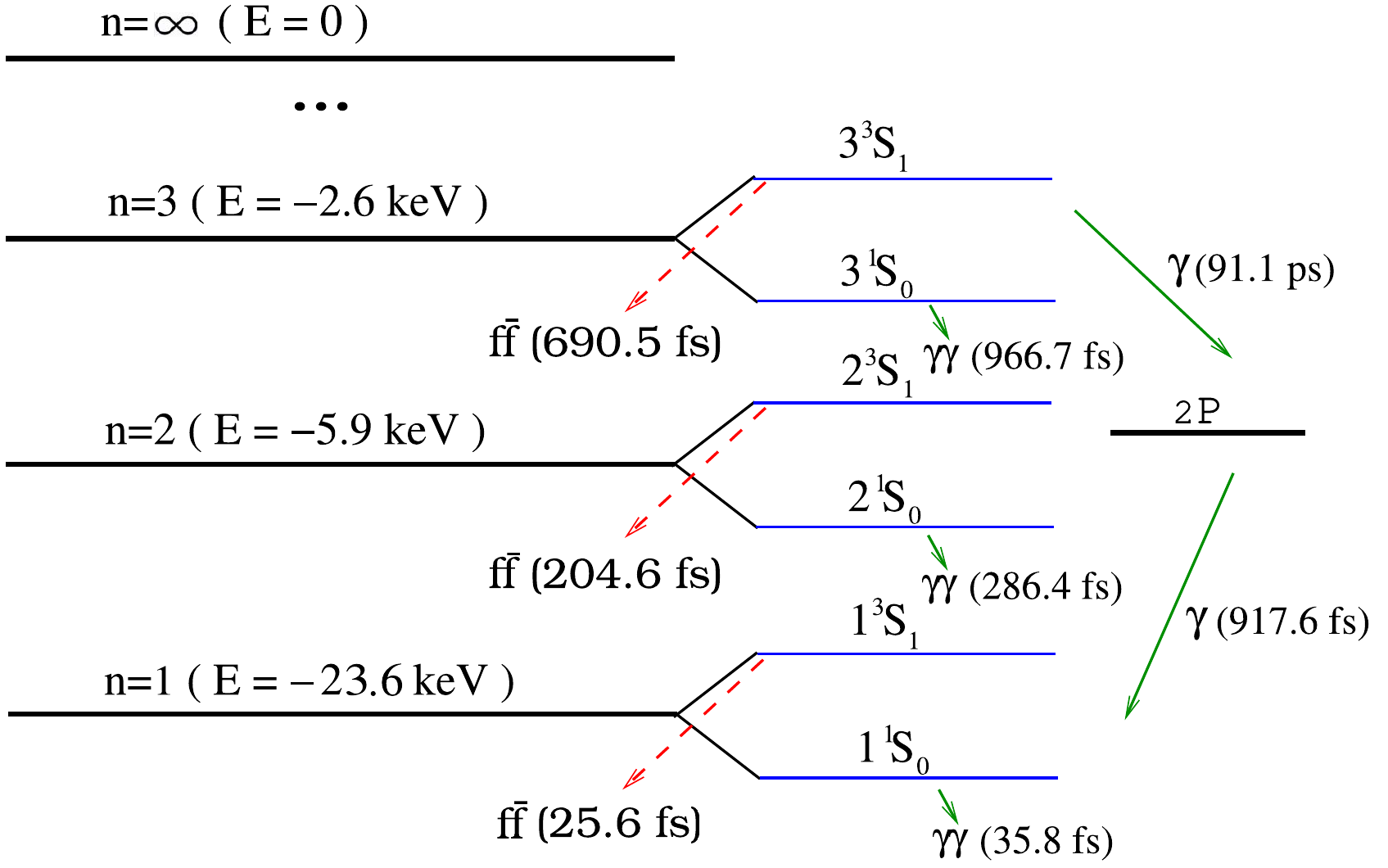}
\caption{Leading-order energy levels and lifetimes of the three lowest ($n=1,2,3$) para- ($n^1\mathrm{S}_0$) and ortho- ($n^3\mathrm{S}_1$) ditauonium states decaying into a pair of photons and of lighter charged fermions ($\ffbar = \epem,\,\mumu,\,\qqbar$), respectively. 
\label{fig:LO_levels}}
\end{figure}

Figure~\ref{fig:LO_levels} shows the LO energy levels, determined from Eq.~(\ref{eq:E_n}), and the LO decay lifetimes determined as explained in Section~\ref{sec:decays}, for the three lowest ditauonium states ($n=1,2,3$). The excited spectrum is obtained considering that a $n^{2S+1}\mathrm{S}_1$ ditauonium state can decay via an electric dipole transition, which conserves the spin quantum number, to a $n^{2S+1}P_J$ state with the emission of a photon with energy $E_n \propto \alpha^2 m_{\tau}$. The radiative transitions from the $3^3\mathrm{S}_1$ state to the $2^3$P state, as well as the transition from the latter to the $1^3\mathrm{S}_1$ state, have energies of
\begin{equation}
E_{n'\to n} = \frac{\alpha^2 m_{\tau}}{4}\left(n^{-2}-n'^{-2}\right) = 
\begin{cases}
      \; -3.28~\text{keV}, & \text{for the } n = 3\to 2\,\text{ transitions},\\
      \; -17.74~\text{keV}, & \text{for the } n = 2\to 1\,\text{ transitions}.
    \end{cases}
\label{eq:E_n_transitions}
\end{equation}
Namely, the Lyman-$\alpha$ photon line of a ditauonium atom transitioning between the first excited ($n = 2$) and the ground ($n = 1$) states has an energy of 17.74~keV.

\subsection{Higher-order corrections}

Equation~(\ref{eq:E_n}) predicts that all ditauonium states sharing the same principal quantum number $n$ are degenerate in energy. However, two types of higher-order corrections are known to break such a degeneracy for QED bound states:
\begin{itemize}
    \item The Lamb shift, due to loop quantum corrections of the Coulomb potential, leads to energy shifts among different $L$ states (S, P, D,...) with the same principal quantum number $n$.
    \item The hyperfine splittings (hfs) ---due to relativistic corrections, the (spin-spin) interaction of the magnetic moments of the components of the bound system, and spin-orbit interactions--- lead to separations between singlet and triplet states.
\end{itemize}

\subsubsection{Lamb shift}

The Lamb shift is an energy splitting of order $\mathcal{O}(\alpha^3)$, \ie\ it is an NLO correction to the leading-order energy spectrum given by Eq.~(\ref{eq:E_n}), and can be computed via the expression
\begin{eqnarray}
      \Delta E_{\rm Lamb}\!&\!=\!&\!\int{d^3\vec{r}\varphi^*(\vec{r})V_U(r)\varphi(\vec{r})},
\end{eqnarray}
where $\varphi(\vec{r})$ is the ditauonium wavefunction, 
and $V_U(r)$ is the nonrelativistic Uehling's potential characterizing difermion insertions into the Coulomb's photon propagator, which is discussed in detail later around Eq.~(\ref{eq:uehling}). Numerically, accounting for the three lepton and quark loop contributions, the energy corrections to the first levels read
\begin{eqnarray}
   \Delta E_{\rm Lamb}^{1\mathrm{S}}\!&\!=\!&\!-115.4~{\rm eV}, \quad \Delta E_{\rm Lamb}^{2\mathrm{S}}=-14.4~{\rm eV},\quad \Delta E_{\rm Lamb}^{2\mathrm{P}}=-8.67~{\rm eV},\nonumber\\
   \Delta E_{\rm Lamb}^{3\mathrm{S}}\!&\!=\!&\!-4.03~{\rm eV},\quad\; \Delta E_{\rm Lamb}^{3\mathrm{P}}=-2.25~{\rm eV},\quad \Delta E_{\rm Lamb}^{3\mathrm{D}}=-1.10~{\rm eV}.
   \label{eq:Lamb_shifts}
\end{eqnarray}
For the lowest 1S ditauonium states, we see that the Lamb effect leads to a $\sim$115~eV downwards shift of their energy levels. Namely, the ditauonium ground-state mass changes from $m_{_{\tata}} = 3553.6963\pm 0.2400$~MeV (adding one extra significant digit to Eq.~(\ref{eq:ditauonium_mass}) for visualization of the effect) to $m_{_{\tata}} = 3553.6962\pm 0.2400$~MeV, an effect that is one thousand times smaller than the current uncertainty of the ditauonium ground-level mass (driven, as aforementioned, by the current tau lepton mass precision).
The uncertainties of the numerical values listed in Eq.~(\ref{eq:Lamb_shifts}) from missing NLO corrections to the Lamb shift (two-loop contribution to the Uelhing potential) are of order
$\mathcal{O}(\alpha^4\,m_{\tau})$, and thus a factor of 10--100 smaller than the values quoted. Such NNLO corrections to the energy levels only impact the accuracy of the energy shifts between
states of different $L$, but not the splitting of different $J$, $S$ states. In the next section, we consider NNLO corrections to the ditauonium energy spectrum that lead to hyperfine splittings
among energy levels with different $J$, $S$ quantum numbers.

\subsubsection{Hyperfine splitting structure}

\begin{figure}[htpb!]
\centering
\includegraphics[width=0.75\textwidth]{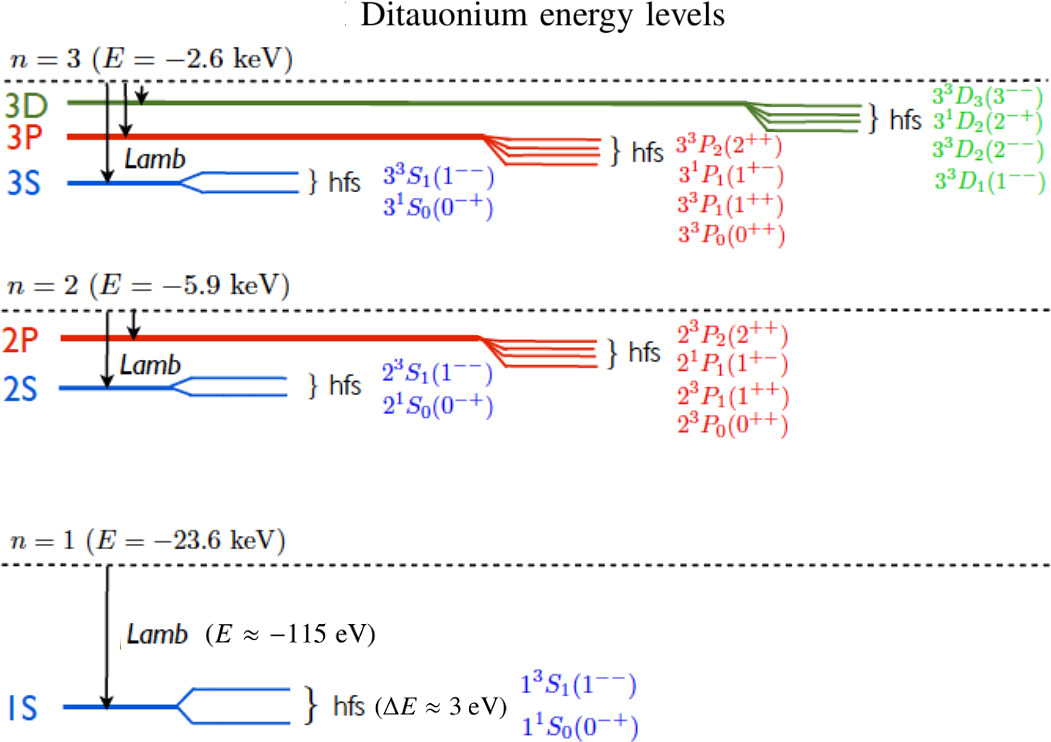}
\caption{Schematic representation of the ditauonium mass spectrum with the effects of the Lamb shift and leading hyperfine splitting contributions shown. For each state $n^{2S+1}L_J$, we also list their associated $(J^{PC})$ quantum numbers. The sizes of each individual Lamb and hyperfine splitting are listed in Eqs.~(\ref{eq:Lamb_shifts}) and ~(\protect\ref{eq:hfs_shifts}), respectively.
\label{fig:massspectrum}}
\end{figure}

The hyperfine structure of a bound state describes the splitting of the singlet and triplet energy levels, as schematically displayed in Fig.~\ref{fig:LO_decays_diags} with the separated $n^1\mathrm{S}_0$ and $n^3\mathrm{S}_1$ states. 
At its lowest-order, the effect is proportional to $\mathcal{O}(\alpha^4\,m_{\tau})$, \ie\ it is an NNLO correction to the leading-order energy spectrum given by Eq.~(\ref{eq:E_n}). The most generic theoretical expression for the hfs QED corrections to true tauonium can be written as an expansion in powers of $(\alpha/\pi)$ and $\alpha\ln(1/\alpha)$, as usually done for the positronium case,
\begin{equation}
\label{eq:E_hfs}
\Delta E_\text{hfs} = \alpha^4 m_{\tau}\left[c_{00} + c_{10} \left(\frac{\alpha}{\pi}\right)  + c_{20}\left(\frac{\alpha}{\pi}\right)^2 + c_{21}\alpha^2\ln(1/\alpha) + c_{30}\left(\frac{\alpha}{\pi}\right)^3 + c_{31}\left(\frac{\alpha}{\pi}\right)^3\ln(1/\alpha) + c_{32}\left(\frac{\alpha}{\pi}\right)^3\ln^2(1/\alpha)  \right],
\end{equation}
where $c_{ij}$ indicates the coefficient of the term proportional to the $\alpha^i \ln^j(1/\alpha)$ correction. All dependencies of the hfs corrections to mass scales other than $m_{\tau}$ are incorporated into the $c_{ij}$ coefficients.
The lowest-order coefficient $c_{00}$ can be determined for any state $n^{2S+1}L_J$ with discrete parity $P=(-)^{L+1}$ and charge $C=(-)^{L+S}$ quantum numbers, via the following generic expression~\cite{Itzykson:1980rh,Berestetskii:1982qgu},
\begin{eqnarray}
c_{00}(n^{2S+1}L_J)\!&\!=\!&\!\frac{11}{64n^4}+\frac{\delta_{S1}}{n^3}\left(\frac{7}{12}\delta_{L0}+\frac{1-\delta_{L0}}{4(2L+1)}B_{JL}\right)-\frac{1}{2n^3(2L+1)},\;\mbox{with}\; B_{JL}=\left\{\begin{array}{lcl}\frac{3L+4}{(L+1)(2L+3)}& {\rm for} & J=L+1,\\
      -\frac{1}{L(L+1)} & {\rm for} & J=L,\\
      -\frac{3L-1}{L(2L-1)} & {\rm for} & J=L-1,\\\end{array}\right.
\end{eqnarray}
where $\delta_{ij}$ is the delta Kronecker symbol.
Using this formula, the leading hfs energy shifts for the S and P ditauonium states amount numerically to:
\begin{eqnarray}
\Delta E_\text{hfs}(1^{1}\mathrm{S}_0)\!&\!=\!&\!-1.653~{\rm eV},\; \Delta E_\text{hfs}(1^{3}\mathrm{S}_1)=+1.286~{\rm eV},\; \Delta E_\text{hfs}(2^{1}\mathrm{S}_0)=-0.261~{\rm eV},\; \Delta E_\text{hfs}(2^{3}\mathrm{S}_1)=+0.107~{\rm eV},\nonumber\\
\Delta E_\text{hfs}(2^{1}\mathrm{P}_1)\!&\!=\!&\!-0.051~{\rm eV},\; \Delta E_\text{hfs}(2^{3}\mathrm{P}_0)=-0.156~{\rm eV},\; \Delta E_\text{hfs}(2^{3}\mathrm{P}_1)=-0.077~{\rm eV},\; \Delta E_\text{hfs}(2^{3}\mathrm{P}_2)=-0.014~{\rm eV},\nonumber\\
\Delta E_\text{hfs}(3^{1}\mathrm{S}_0)\!&\!=\!&\!-0.083~{\rm eV},\; \Delta E_\text{hfs}(3^{3}\mathrm{S}_1)=+0.026~{\rm eV},\;
\Delta E_\text{hfs}(3^{1}\mathrm{P}_1)=-0.020~{\rm eV},\; \Delta E_\text{hfs}(3^{3}\mathrm{P}_0)=-0.052~{\rm eV},\;\label{eq:hfs_shifts}\\ 
\Delta E_\text{hfs}(3^{3}\mathrm{P}_1)\!&\!=\!&\!-0.028~{\rm eV},\; \Delta E_\text{hfs}(3^{3}\mathrm{P}_2)=-0.0095~{\rm eV}\nonumber.
\end{eqnarray}
From these results, one can see, first, that the energy splittings of the S-wave family are positive (negative) for the $n^3$S ($n^1$S) states, whereas the mass hierarchy in the P waves start with $n^3\mathrm{P}_2$, followed by $n^1\mathrm{P}_1$ and $n^3\mathrm{P}_1$, with the lightest state being $n^3\mathrm{P}_0$. Figure~\ref{fig:massspectrum} displays the ditauonium mass spectrum including the higher-order effects due to Lamb shifts, Eq.~(\ref{eq:Lamb_shifts}), and leading contributions to the hfs structure, Eq.~(\ref{eq:hfs_shifts}), computed here.

Calculations exist for the higher-order hfs $c_{ij}$ coefficients of Eq.~(\ref{eq:E_hfs}) for $(i,j)\neq (0,0)$, in the case of positronium and dimuonium~\cite{Karplus:1952wp,Lepage:1977gd,Czarnecki:1998zv}. Such higher-order $c_{ij}$ coefficients are of order unity but suppressed by increasing powers of $(\alpha)^i \ln^j(1/\alpha)$, and therefore they are numerically in the sub-eV range. Given the short lifetimes of the ditauonium systems and the difficulty to produce them in very large quantities for experimental study~\cite{dEnterria:2022ysg,DdEHSS}, it is unlikely that one will ever be able to carry out such accurate spectroscopic studies of their higher-order hyperfine splitting structure. For this reason, we just retain the first term of Eq.~(\ref{eq:E_hfs}), $\Delta E_\text{hfs}=\alpha^4\,m_\tau\cdot c_{00}$, that indicates an ortho- minus para-state energy difference of $\Delta E_\text{hfs}^{3\mathrm{S}-1\mathrm{S}} =  1.286 + 1.653 \approx 2.94$~eV for the ditauonium ground states.

\section{Decay widths, branching fractions, and lifetimes}
\label{sec:decays}

Two main differences separate ditauonium from its lightest relatives positronium and dimuonium: the much larger mass and much shorter lifetime of its constituents leptons. First, the significantly more massive ditauonium system has accessible decays to lighter fermion pairs that are kinematically forbidden in the positronium and dimuonium cases. Indeed, ortho-ditauonium can decay into five different pairs of standard model (SM) charged fermions\footnote{The $\ccbar$ decay is not kinematically accessible because the lightest charm (D) mesons have masses that are more than half the ditauonium mass. It is however possible for ditauonium to decay into light charmonium states plus a photon (such as $(\tau^+\tau^-)_0 \to J/\psi+\gamma$ and $(\tau^+\tau^-)_1 \to \eta_c+\gamma$)~\cite{DdEHSS}. However, such modes are suppressed by a $\mathcal{O}(\alpha v_c^3)$ factor (where $v_c$ is the relatively low velocity of the bound charm quarks) and a small phase space compared to the leading decay channels, and are therefore ignored here.}, $(\tautau)_1\to\ffbar$ (with $f = e, \mu, u, d, s$) via the annihilation of both constituent $\tau$ leptons into an intermediate single photon (Fig.~\ref{fig:LO_decays_diags}, right), whereas ortho-positronium can decay to none (except to $\nunubar$ with negligible branching ratio~\cite{Govaerts:1996dt,Czarnecki:1999mt}, if one considers all fermions and not just the charged ones, see Section~\ref{sec:nunu}), and ortho-dimuonium can only decay to $\epem$ pairs. 
Similarly, para-ditauonium has higher-order $(\tautau)_0\to\gamma\ffbar$ Dalitz decays that are not possible in the positronium and dimuonium cases. This opens up multiple different decay partial widths for the ditauonium states, which increase by a few percent their total width compared to just considering their leading diphoton and difermion decays as we show below. Second, whereas positronium is constituted by a pair of $\epem$ that are each one of them individually stable, and whereas the $\mu^\pm$ in dimuonium have very large lifetimes ($\tauup = 2.197~\mu$s) for particle physics standards, the $\tau$ leptons composing ditauonium are short-lived objects ($\tauup = 290.3$~fs) that can individually decay weakly before their bound system itself does. When any of the two tau leptons decays through the weak interaction, the bound state disappears with an effective decay width\footnote{Numerical conversion of lifetimes and widths in natural units is done via: $\Gamma~(\text{eV}) = 0.658212/\tauup~(\text{fs})$.} of
\begin{equation}
\Gamma_{(2)\tau\to X} = 2/\tauup = 0.004535 \pm 0.000008~\mathrm{eV}.
\label{eq:taudecaywidth}
\end{equation}
As we show below, this leaves only the two lowest ditauonium states as those that can form actual bound states before their inner leptons decay freely.

The subsections below present the leading- and higher-order QED corrections for the para- and ortho-ditauonium decays. Electroweak corrections are comparatively suppressed by an $\mathcal{O}(\alpha^7 m_\tau)$ factor~\cite{Lamm:2015lia}, and neglected here. Higher quantum-chromodynamics (QCD) corrections impact decay modes involving quark-antiquark pairs, and they are either phenomenologically incorporated into the computed ortho-ditauonium decays through the experimentally measured $R_\text{had}$ ratio of hadronic to dimuon cross sections in $\epem$ collisions, or suppressed by at least a factor of $\alpha\alphas$ with respect to the leading decay width in the para-ditauonium case, and also neglected hereafter.

\subsection{Leading-order results}

As shown in Fig.~\ref{fig:LO_decays_diags} (left and center), the dominant para-ditauonium decay is into two photons. 
At zeroth order, the diphoton decay rate of a 
spin-singlet bound state of principal quantum number $n$ is given by~\cite{Itzykson:1980rh,Berestetskii:1982qgu}
\begin{equation}
\Gamma^{(0)}(n {}^1\mathrm{S}_0\to\gaga) = \frac{\alpha^5 \, m_\tau}{2\,n^3} = 
\begin{cases}
      \; 0.018384~\text{eV}, & \text{for the } n=1\,\text{ state}\\
      \; 0.002298~\text{eV}, & \text{for the } n=2\,\text{ state}\\
      \; 0.000681~\text{eV}, & \text{for the } n=3\,\text{ state}
\end{cases}.
\label{eq:gaga_tata_width}
\end{equation}

Ortho-ditauonium decays preferentially into all pairs of SM fermions lighter than half the ditauonium mass, ${1^3\mathrm{S}_1}\to\ffbar$ ($f = e, \mu, u, d, s$), through $\tautau$ annihilation into an intermediate single photon (Fig.~\ref{fig:LO_decays_diags}, right) with LO partial widths,
\begin{eqnarray}
\Gamma^{(0)}(n {}^3\mathrm{S}_1 \to \ffbar)\!&\!=\!&\!N_{c,f} Q_f^2\,\frac{\alpha^5\, m_\tau}{6\, n^3}\left(1+\frac{2m_f^2}{m_{_{\tata}}^2}\right)\sqrt{1-\frac{m_f^2}{m_\tau^2}},
\label{eq:LO_orthodecaywidth}
\end{eqnarray}
where $Q_f=-1$ for charged leptons and $Q_f=\frac{2}{3},-\frac{1}{3},-\frac{1}{3}$ for $u, d, s$ quarks, respectively, and $N_{c,f}$ is the number of colours of the fermion, \ie\ $N_{c,f}=1$ for leptons and $N_{c,f}=N_c=3$ for quarks.
Given the smallness of the electron and muon masses compared to the tau one, the product of the two last terms in Eq.~(\ref{eq:LO_orthodecaywidth}) is basically unity for all numerical purposes, and the zeroth-order dilepton decay width of ortho-ditauonium reads
\begin{align}
\Gamma^{(0)}(n {}^3\mathrm{S}_1 \to\epem,\mumu) & = \frac{\alpha^5 m_{\tau}}{6 n^3} = 
\begin{cases}
      \; 0.006128~\text{eV}, & \text{for the } n=1\,\text{ state}\\
      \; 0.000766~\text{eV}, & \text{for the } n=2\,\text{ state}\\
      \; 0.000227~\text{eV}, & \text{for the } n=3\,\text{ state}
\end{cases}.
\end{align}
with decays into $\epem$ and $\mumu$ basically happening at the same rate. 
The values of the inclusive quark-pair decay partial widths can be numerically derived from the same dilepton expression above multiplied by the experimentally measured value of the ratio of hadronic to dimuon cross sections, $R_\text{had}\equiv\sigma(\epem\to\gamma^*\to\qqbar)/\sigma(\epem\to\gamma^*\to\mumu)\approx \sum_{q=u,d,s} N_c Q_q^2$ evaluated at the ditauonium mass scale. Namely,
\begin{equation}
\Gamma^{(0)}(n {}^3\mathrm{S}_1 \to\qqbar) = \frac{\alpha^5 m_{\tau}}{6 n^3}R_\text{had}(m_{_{\tata}}^2)\, \approx 2.2\frac{\alpha^5 m_{\tau}}{6 n^3} = 
\begin{cases}
      \; 0.013482~\text{eV}, & \text{for the } n=1\,\text{ state}\\
      \; 0.001685~\text{eV}, & \text{for the } n=2\,\text{ state}\\
      \; 0.000499~\text{eV}, & \text{for the } n=3\,\text{ state}
\end{cases},.\label{eq:LOwidthorthohad}
\end{equation}
where the last approximate numerical equality uses the current experimental value of the $R_\text{had}$ ratio (Table~\ref{tab:PDG}).

The total LO width of ortho-ditauonium ---adding dielectron, dimuon, and diquark decays--- amounts thus to:
\begin{equation}
\Gamma^{(0)}(n {}^3\mathrm{S}_1) = 4.2\frac{\alpha^5 m_{\tau}}{6 n^3} = 
\begin{cases}
      \; 0.025738~\text{eV}, & \text{for the } n=1\,\text{ state}\\
      \; 0.003217~\text{eV}, & \text{for the } n=2\,\text{ state}\\
      \; 0.000953~\text{eV}, & \text{for the } n=3\,\text{ state}.
\end{cases}.\label{eq:LOwidthortho}
\end{equation}

For excited states with $L > 0$, the overlap of the wavefunctions is zero resulting in preferential radiative decays to lower 1S states rather than annihilation decays.
The radiative decay widths of the excited (3S and 2P) states can be derived from the generic expressions for the ratios of decay widths of true leptonium states,
\begin{align*}
\frac{\Gamma^{(0)}(n {}^1\mathrm{S}_0 \to\gaga)}{\Gamma^{(0)}(2\mathrm{P} \to 1\mathrm{S})} & =
\left(\frac{3}{2} \right)^8 \frac{1}{n^3} = \frac{25.6}{n^3} , &
\frac{\Gamma^{(0)}(2\mathrm{P} \to 1\mathrm{S})}{\Gamma^{(0)}(3\mathrm{S} \to 2\mathrm{P})} & = \left( \frac{5}{3} \right)^9 = 99.2\,.
\end{align*}
The partial decays widths from higher to lower orbitals are thus given by
\begin{align}
\Gamma^{(0)}(2\mathrm{P} \to 1\mathrm{S}) & = \left(\frac{2}{3}\right)^8 \frac{\alpha^5 m_{\tau}}{2} , & 
\Gamma^{(0)}(3\mathrm{S} \to 2\mathrm{P}) & = \left(\frac{2}{5}\right)^9 \frac{3\alpha^5 m_{\tau}}{4}.
\end{align}
The annihilation partial decay widths of P-wave states are at least of order $\mathcal{O}(\alpha^7)$ and thus comparatively negligible.

The zeroth-order results for the energy levels and leading decays of ditauonium states are shown in Fig.~\ref{fig:LO_levels}, where the lifetime has been derived from each width via $\tauup = 1/\Gamma$. From this plot, we see first that only the two lowest-energy levels, para-ditauonium $1^1\mathrm{S}_0$ (with $\tauup = 35.8$~fs) and
ortho-ditauonium $1^3\mathrm{S}_1$ (with $\tauup = 25.6$~fs) have lifetimes smaller than half that of its constituent tau leptons ($\tauup = 145.15$~fs), and can thereby form a transient bound state before the weak decay of any of them. Second, one can observe that the lifetimes of para- and ortho-states are relatively similar for ditauonium, at variance with the positronium case where the ortho-state can only kinematically decay into the 3-photon mode (which is, at least, $\alpha$ times suppressed compared to the para-positronium 2-photon decay) and thereby features three orders-of-magnitude longer lifetime than the para-state.

\subsection{Para-ditauonium decay}

\subsubsection{Virtual NLO and (partial) NNLO corrections}

\begin{figure}[htpb!]
\centering
\includegraphics[width=0.99\textwidth]{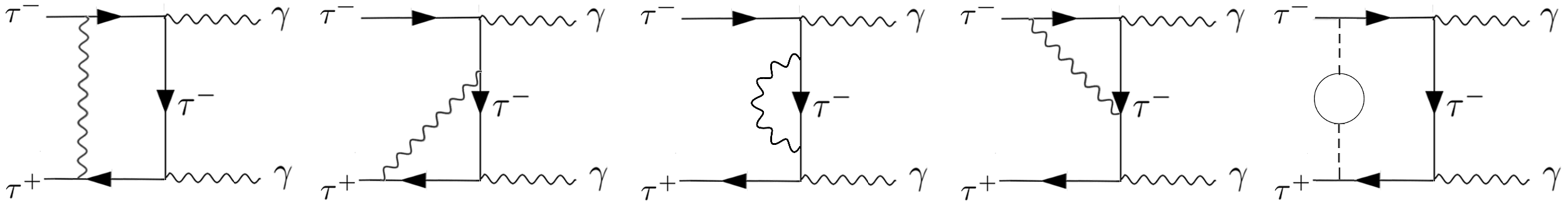}
\caption{Representative diagrams of NLO virtual corrections of the para-ditauonium diphoton width. The first four diagrams show radiative virtual QED corrections, and the rightmost diagram represents the modifications 
of the ditauonium wavefunction at the origin due to vacuum polarization insertions (circle) into the Coulomb's photon propagator (dashed line).
\label{fig:NLO_virt_para_diags}}
\end{figure}

Figure~\ref{fig:NLO_virt_para_diags} shows the NLO virtual corrections to the diphoton para-ditauonium partial width, $\Gamma(n^1\mathrm{S}_0\to\gaga)$. They include virtual photon exchanges between the tau leptons, as well as modifications of the ditauonium wavefunction at the origin, $|\varphi_{n\mathrm{S}}(0)|$, due to vacuum polarization effects affecting the common Coulomb potential of the two tau leptons and that are effectively of order $\mathcal{O}(\alpha/\pi)$. These latter effects are indicated in the rightmost panel of Fig.~\ref{fig:NLO_virt_para_diags} by the fermion loop insertion to the Coulomb potential (dashed line) typical of NRQED calculations~\cite{Kinoshita:1995mt}. Calculations for all these higher-order corrections exist for the dimuonium case~\cite{Malenfant:1987tm,Jentschura:1997tv,Ginzburg:1998df,Jentschura:1998vkm,Lamm:2015lia,Ji:2016fat}, and are extended and applied to ditauonium in this section.

The radiative virtual QED corrections are given by the set of four leftmost diagrams displayed in Fig.~\ref{fig:NLO_virt_para_diags}. These virtual NLO corrections were exactly calculated in~\cite{Brown:1952eu,Harris:1957zza} for para-positronium. In the nonrelativistic limit, the overall sum of such diagrams amounts to a relative width correction of size
\begin{equation}\label{eq:NLO_Vert-2A}
\Delta\Gamma^\text{NLO}_{\text{virt.exch}}(n^1\mathrm{S}_0\to\gaga)=C^\text{NLO}_{\text{virt.exch}}\left(\frac{\alpha}{\pi}\right)\Gamma^{(0)}(n ^1\mathrm{S}_0\to\gaga),\; \mbox{ with }\; C^\text{NLO}_{\text{virt.exch}} = -\frac{20-\pi^2}{4} = -2.5326.
\end{equation}
The universality of the correction in this limit justifies its use also for the para-dimuonium case~\cite{Jentschura:1997tv,Jentschura:1998vkm}, and hence also for para-ditauonium in the present study. 

The loop in the Coulomb's photon propagator of Fig.~\ref{fig:NLO_virt_para_diags}  (right) represents modifications of order $\mathcal{O}(\alpha/\pi)$ of the wavefunction at the origin.
For para-ditauonium, the corrections incorporate all fermion loops, though they are dominated by the contribution of the lightest electron loop. The first approach to obtain this correction, according to the expression
\begin{equation}
\Delta\Gamma^\text{NLO}_{\text{Coul}}(n ^1\mathrm{S}_0\to\gaga)=\frac{\Delta|\varphi_{n\mathrm{S}}(0)|^2}{|\varphi_{n\mathrm{S}}(0)|^2}\Gamma^{(0)}(n ^1\mathrm{S}_0\to\gaga)=C_\text{Coul}^{\text{NLO},n\mathrm{S}}\cdot \left(\frac{\alpha}{\pi}\right)\cdot \Gamma^{(0)}(n ^1\mathrm{S}_0\to\gaga),
\label{eq:NLO_Coul}
\end{equation}
was performed in~\cite{Malenfant:1987tm} and further improved in~\cite{Jentschura:1997tv,Jentschura:1998vkm}. This NLO coefficient for the $n$S para-ditauonium states can be computed via the integral 
\begin{equation}\label{eq:CVP}
C_\text{Coul}^{\text{NLO},n\mathrm{S}}\cdot\left(\frac{\alpha}{\pi}\right)=-2a_0^3\int_0^{+\infty} d\varrho\varrho^2K_{n\mathrm{S}}(\varrho)\overline{g}_{n\mathrm{S}}(\varrho)\,V_U(a_0\varrho),
\end{equation}
where, first, $\overline{g}_{n\mathrm{S}}(\varrho)$ are the reduced Green's functions of the dimensionless radial variable $\varrho=r/a_0$ for the nonrelativistic Coulomb potential, given by the following expressions for the 1S and 2S states, respectively,
\begin{eqnarray}\label{eq:green1}
\overline{g}_{1\mathrm{S}}(\varrho)\!&\!=\!&\!-\alpha m_\text{red}^2\frac{e^{-\varrho}}{\varrho}\left[4\varrho(\ln2\varrho+\gamma_E)+4\varrho^2-10\varrho-2\right],\\
\overline{g}_{2\mathrm{S}}(\varrho)\!&\!=\!&\!\alpha m_\text{red}^2\frac{e^{-\varrho/2}}{2\varrho}\left[4\varrho(\varrho-2)(\ln\varrho+\gamma_E)+\varrho^3-13\varrho^2+6\varrho+4\right],
\end{eqnarray}
with $\gamma_E\approx 0.577216$ the Euler's constant, 
and where the integration kernels read
\begin{eqnarray}
K_{1\mathrm{S}}(\varrho)\!&\!=\!&\!e^{-\varrho},\quad K_{2\mathrm{S}}(\varrho)=e^{-\varrho/2}(1-\varrho/2).
\end{eqnarray}
Secondly, the $V_U(r)$ function in Eq.~(\ref{eq:CVP}) is the nonrelativistic Uehling's potential, characterizing polarization insertions into the Coulomb's photon propagator~\cite{Jentschura:1997tv,Jentschura:1998vkm}, given by
\begin{equation}\label{eq:uehling}
V_U(r)=-\sum_{f}{N_{c,f}Q_f^2\left(\frac{\alpha}{\pi}\right)\int_0^1dv_f\frac{v_f^2(1-v_f^2/3)}{1-v_f^2}\frac{\alpha e^{-\lambda_{f} r}}{r}}, 
\end{equation}
where $v_f$ is the integration variable defined as $v_f^2=1-4m_f^2/s$, with $s$ the square of the invariant mass of the fermion pair in the loop, and where the effective Coulomb's photon mass $\lambda_{f}$ (with dimension $L^{-1}$) takes the values,
\begin{equation}\label{eq:lambdas}
    \lambda_f=\frac{2m_f}{\sqrt{1-v_f^2}},
\end{equation}
upon insertion of a fermion $f$ loop. Such a potential is known fully analytically expressed in terms of modified Bessel functions and Bickley functions~\cite{Frolov:2011rm}. It is worth mentioning that the integral's kernel term proportional to $v_f^2(1-v_f^2/3)/(1-v_f^2)$ does not depend on $m_f$ and so do not either the integration bounds. Hence, the mass dependence on the type of loop mainly arises from $\lambda_{f}$ in the argument of the exponential term. 
After cross-checking that with Eq.~(\ref{eq:CVP}) we can reproduce the same values of the NLO coefficients derived for the para-dimuonium case in refs.~\cite{Jentschura:1997tv,Jentschura:1998vkm}, we determine the numerical coefficients
\begin{equation}
    C_\text{Coul}^{\text{NLO},1\mathrm{S}} =5.805,\; \mbox{ and }\; C^\text{NLO,2S}_\text{Coul}=4.429,
    \label{eq:bound_green}
\end{equation}
for S-wave ditauonium by inserting in the loop all fermions $f=\{e,\mu,\tau,u,d,s\}$ with the masses listed in Table~\ref{tab:PDG}. These Coulomb corrections are clearly dominated by the electron loop (accounting for about 90\% of the total value) and, therefore, the relatively large uncertainties of the constituent masses of the inserted light quarks are not numerically important.

In addition, virtual NNLO corrections related to the low energy logarithmic contribution arising from spin-spin contact interactions and pure orbit corrections in the Breit Hamiltonian~\cite{Berestetskii:1982qgu}, which were first derived for para-positronium~\cite{Caswell:1978cz,Khriplovich:1990eh}, are found to amount to:
\begin{equation}\label{eq:NNLO_para_virt}
\Delta\Gamma^\text{NNLO}_\text{Breit}(1^1\mathrm{S}_0\to\gaga) = C^\mathrm{NNLO}_\text{Breit} \left(\frac{\alpha}{\pi}\right)^2\Gamma^{(0)}(1^1\mathrm{S}_0\to\gaga),\quad \text{with}\quad
C^\mathrm{NNLO}_\text{Breit}=2\pi^2\ln (1/\alpha) = 97.1217,
\end{equation}
for the $n=1$ (valid also for $n=2$) para-ditauonium case.

\subsubsection{Real NLO and (partial) NNLO corrections}

\begin{figure}[htpb!]
\centering
\includegraphics[width=0.99\textwidth]{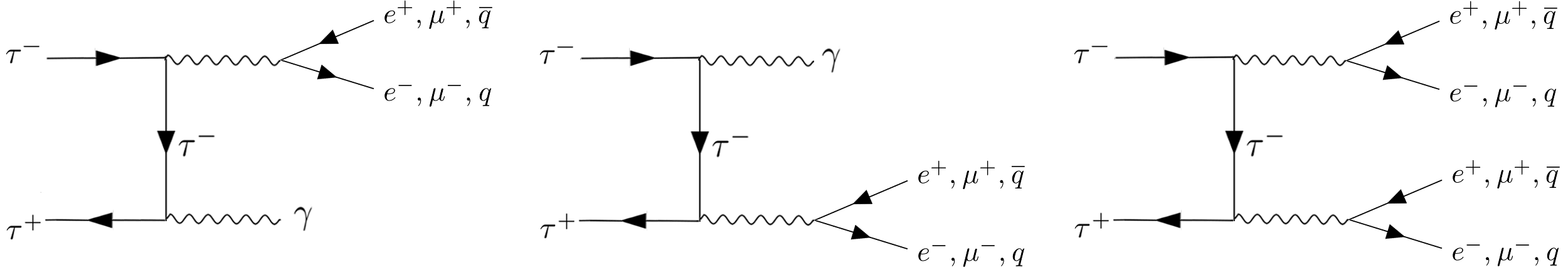}
\caption{Left and center: Diagrams for Dalitz decays of para-ditauonium (NLO real corrections to the diphoton $n^1\mathrm{S}_0$ width). Right: ``Double Dalitz'' (4-fermions) NNLO real correction of the para-ditauonium decay.
\label{fig:Dalitz_decays_diags}}
\end{figure}

The higher-order real corrections to para-ditauonium decays include, first, the splitting of one of the decay photons into any kinematically allowed fermion-antifermion pair (dielectron, dimuon, and light diquarks). The corresponding diagrams are shown in the left and center panels of Fig.~\ref{fig:Dalitz_decays_diags}. Beyond this, the rightmost panel of Fig.~\ref{fig:Dalitz_decays_diags} shows the NNLO double photon splitting into four fermions.

The Dalitz $\gamma \ffbar$ decay width can be calculated at the lowest order as
\begin{equation}
\Gamma^\text{NLO}_\text{Dalitz}(n^1\mathrm{S}_0\to\gamma\ffbar)= N_{c,f} Q_f^2\frac{2\alpha^6 m_\tau}{9\pi n^3}\left[3\ln{\left(\frac{m_\tau}{m_f}+\sqrt{\frac{m_\tau^2}{m_f^2}-1}\right)}-\left(4-\frac{m_f^2}{m_\tau^2}\right)\sqrt{1-\frac{m_f^2}{m_\tau^2}}\right].
\end{equation}
Such a perturbative formula applies well for the $f=e,\mu$ leptons, for which we find
\begin{eqnarray}
\hspace{-5.5mm}\Gamma^\text{NLO}_\text{Dalitz}(n^1\mathrm{S}_0\to\gamma \epem) \!&\!=\!&\! 22.5414\,\frac{2\alpha^6 m_\tau}{9\pi n^3} = C^{\rm NLO}_{\rm Dalitz}(\epem)\,\left(\frac{\alpha}{\pi}\right)\cdot \Gamma^{(0)}(n^1\mathrm{S}_0\to\gaga),\; \mbox{ with }\; C^{\rm NLO}_{\rm Dalitz}(\epem)=10.0184, \label{eq:log_mass_el}\\
\hspace{-5mm}\Gamma^\text{NLO}_\text{Dalitz}(n^1\mathrm{S}_0\to\gamma \mumu) \!&\!=\!&\! 6.55457\,\frac{2\alpha^6 m_\tau}{9\pi n^3} = C^{\rm NLO}_{\rm Dalitz}(\mumu)\,\left(\frac{\alpha}{\pi}\right)\cdot \Gamma^{(0)}(n^1\mathrm{S}_0\to\gaga),\; \mbox{ with }\; C^{\rm NLO}_{\rm Dalitz}(\mumu)=2.9131. \label{eq:log_mass_mu}
\end{eqnarray}
For the quarks, we need in addition to sum over the three quark flavour charges as follows,
\begin{eqnarray}
\Gamma^\text{NLO}_\text{Dalitz}(n^1\mathrm{S}_0\to\gamma\qqbar)\!&\!=\!&\!\frac{2\alpha^6 m_\tau}{9\pi n^3}N_c\sum_{q}{Q_q^2\left[3\ln{\left(\frac{m_\tau}{m_q}+\sqrt{\frac{m_\tau^2}{m_q^2}-1}\right)}-\left(4-\frac{m_q^2}{m_\tau^2}\right)\sqrt{1-\frac{m_q^2}{m_\tau^2}}\right]}\nonumber\\
\!&\!\approx\!&\!\frac{2\alpha^6 m_\tau}{9\pi n^3}\left[\frac{9\pi}{2} \frac{\Delta\alpha_\text{had}^{(3)}(m_{_{\tata}}^2)}{\alpha}-\frac{3}{2}R_\text{had}(m_\tau^2)\right]=11.6\,\frac{2\alpha^6 m_\tau}{9\pi n^3}\nonumber\\
\!&\!=\!& C^{\rm NLO}_{\rm Dalitz}(\qqbar)\,\left(\frac{\alpha}{\pi}\right)\cdot \Gamma^{(0)}(n^1\mathrm{S}_0\to\gaga)\; \mbox{ with }\; C^{\rm NLO}_{\rm Dalitz}(\qqbar)=5.16.
\label{eq:log_mass_qqbar}
\end{eqnarray}
In the equation above, we have replaced the purely perturbative expression with quark masses of the first line with the second-line expression that employs the hadronic quantities $\Delta\alpha_\text{had}^{(3)}(m_{_{\tata}}^2)$ and $R_\text{had}(m_\tau^2)$ listed in Table~\ref{tab:PDG}, where the hadronic running of the QED coupling, $\Delta\alpha_\text{had}^{(3)}(m_{_{\tata}}^2)$, can be obtained from the experimental $R_\text{had}(s)$ ratio via dispersion relations.

Finally, we compute the NNLO double-real correction given by the ``double Dalitz'' diagram shown in Fig.~\ref{fig:Dalitz_decays_diags} (right). The partial widths can be cast into the following generic form:
\begin{equation}\label{eq:doubleDalitz}
\Gamma^\text{NNLO}_\text{2Dalitz}(n^1\mathrm{S}_0\to\ffbar\ffbarprime) = C^\mathrm{NNLO}_\text{2Dalitz}(\ffbar\ffbarprime) \left(\frac{\alpha}{\pi}\right)^2\Gamma^{(0)}(n^1\mathrm{S}_0\to\gaga)\,,
\end{equation}
with the values of $C^\mathrm{NNLO}_\text{2Dalitz}(\ffbar\ffbarprime)$ calculated numerically with the phase space integrator of {\sc\small HELAC-Onia}~\cite{Shao:2012iz,Shao:2015vga}, and amounting to
\begin{eqnarray}
      C^\mathrm{NNLO}_\text{2Dalitz}(\epem\epem)\!&\!=\!&\!23.40,\quad C^\mathrm{NNLO}_\text{2Dalitz}(\epem\mumu)\!=\!13.87,\quad 
      C^\mathrm{NNLO}_\text{2Dalitz}(\epem\qqbar)\!=\!12.07,\nonumber\\
      C^\mathrm{NNLO}_\text{2Dalitz}(\mumu\mumu)\!&\!=\!&\!1.66,\quad\;\; C^\mathrm{NNLO}_\text{2Dalitz}(\mumu\qqbar)\!=\!2.74,\quad\quad\; C^\mathrm{NNLO}_\text{2Dalitz}(\qqbar\qqbarprime)\!=\!0.83.
      \label{eq:2dalitz_results}
\end{eqnarray}
In the above equation, the notation ``$\qqbar\qqbarprime$'' indicates the sum of all four quark final states, including both same flavour and different flavour channels.
It is interesting to notice that in the asymptotic heavy-$m_\tau$ limit, the coefficients for the \textit{same-flavour} $4$-lepton final states approach the results of the expression
\begin{eqnarray}
      C^\mathrm{NNLO}_\text{2Dalitz}(\lele\lele)\!&\!=\!&\!\frac{4}{9}\ln^2{\left(\frac{m_\tau}{m_\ell}\right)}-\ln{\left(\frac{m_\tau}{m_\ell}\right)}+2+\mathcal{O}\left(\frac{m_\ell}{m_\tau}\right),
\end{eqnarray}
with the double logarithm term matching that given in Refs.~\cite{Caswell:1978cz,Khriplovich:1990eh,Jentschura:1998vkm} for positronium and dimuonium. Similarly, the asymptotic formulas read,
\begin{eqnarray}
      C^\mathrm{NNLO}_\text{2Dalitz}(\qqbar\qqbar)\!&\!=\!\!&\!\!\sum_{q=u,d,s}{Q_q^4\left[\frac{4}{9}N_c^2\ln^2{\left(\frac{m_\tau}{m_q}\right)}-N_c^2\ln{\left(\frac{m_\tau}{m_q}\right)}+2\left(1.0762(N_c^2-N_c)+N_c\right)+\mathcal{O}\left(\frac{m_q}{m_\tau}\right)\right]},
\end{eqnarray}
for the \textit{same-flavour} $4$-quark final states, and,
\begin{eqnarray}
      C^\mathrm{NNLO}_\text{2Dalitz}(\ffbar\ffbarprime)\!&\!=\!&\!Q_{f}^2Q_{f^\prime}^2N_{c,f}N_{c,f^\prime}\left[\frac{4}{9}\ln{\left(\frac{m_\tau}{m_f}\right)}\ln{\left(\frac{m_\tau}{m_{f^\prime}}\right)}+3\left(\ln{\left(\frac{m_\tau}{m_f}\right)}+\ln{\left(\frac{m_\tau}{m_{f^\prime}}\right)}\right)-28.5+\mathcal{O}\left(\frac{m_f}{m_\tau},\frac{m_{f^\prime}}{m_\tau}\right)\right],
\end{eqnarray}
for the generic different-flavour double-Dalitz case.


\subsubsection{Combined higher-order corrections}

The total annihilation decay width of para-ditauonium (we focus just on the $n=1$ state hereafter) can be written in a compact form, by combining all previous results, as follows
\begin{equation}
\Gamma_\text{tot}(1^1\mathrm{S}_0)=\left(1+P_\text{1S,tot}^\text{NLO}\left(\frac{\alpha}{\pi}\right)+P_\text{1S,tot}^\mathrm{NNLO^{\star}}\left(\frac{\alpha}{\pi}\right)^2\right)\Gamma^{(0)}(1^1\mathrm{S}_0).
\label{eq:para_width_corrs}
\end{equation}
The LO total annihilation decay width of para-ditauonium has a single one-channel contribution, i.e.,
\begin{eqnarray}
    \Gamma^{(0)}(1^1\mathrm{S}_0)&=&\Gamma^{(0)}(1^1\mathrm{S}_0\to \gamma\gamma),
\end{eqnarray}
which is given by Eq.~(\ref{eq:gaga_tata_width}). The $P_\text{1S,tot}^\text{NLO}$ parameter accounts for the numerical NLO coefficients derived in Eqs.~(\ref{eq:NLO_Vert-2A}), (\ref{eq:NLO_Coul}), (\ref{eq:bound_green}), (\ref{eq:log_mass_el}), (\ref{eq:log_mass_mu}), 
and (\ref{eq:log_mass_qqbar}), \ie
\begin{equation}
    P_\text{1S,tot}^\text{NLO}= C^\text{NLO}_{\text{virt.exch}}
    +C_\text{Coul}^\text{NLO,1S}+\sum_{f}{C_\text{Dalitz}^\text{NLO}(\ffbar)} = -2.533 + 5.805 + 10.018 + 2.913 + 5.16 = 21.36\,,
    \label{eq:p1tot}
\end{equation}
where one can see that the largest higher-order contribution is from the $\gamma\epem$ Dalitz decay, which accounts for about half of all NLO corrections. $P_\text{1S,tot}^\mathrm{NNLO^{\star}}$ is a numerical coefficient accounting for the partial virtual and real NNLO corrections given, respectively, by Eqs.~(\ref{eq:NNLO_para_virt}), (\ref{eq:doubleDalitz}) and (\ref{eq:2dalitz_results}),
\begin{equation}\label{eq:para_nnlo_tot}
P_\text{1S,tot}^\mathrm{NNLO^{\star}}= C^\mathrm{NNLO}_\text{Breit} + \sum_{f,f^\prime}{C_\text{2\,Dalitz}^\mathrm{NNLO}(\ffbar\ffbarprime)} =  97.12 + 54.57 = 151.69\,.
\end{equation}
The corresponding values of all individual coefficients are listed in Table~\ref{tab:paraditauonium_corrs1}.

\begin{table}[htbp!]
\tabcolsep=1.1mm
\centering
\caption{Coefficients for the sum of all NLO and partial NNLO corrections of each partial para-ditauonium decay channel, $C^\text{NLO,NNLO}_\text{part}$, and final $P^\text{NLO}_\text{1S,tot}$ and $P_\text{1S,tot}^\mathrm{NNLO^{\star}}$ correction terms as defined in Eqs.~(\ref{eq:p1tot}) and (\ref{eq:para_nnlo_tot}) for its total annihilation decay width. 
\label{tab:paraditauonium_corrs1}}
\vspace{0.1cm}
\begin{tabular}{cccccc|ccccccccc}\hline
$C^\text{NLO}_\text{part}$: & $\gaga$ & $\gamma\epem$ & $\gamma\mumu$ & $\gamma\qqbar$ & \hspace{1.5mm} $P^\text{NLO}_\text{1S,tot}$ \hspace{1.5mm}& $C^\text{NNLO}_\text{part}$: & $\gaga$ & $2\epem$ & $\epem\mumu$ & $\epem\qqbar$ & $2\mumu$ & $\mumu\qqbar$ & $\qqbar\qqbarprime$ &  \hspace{1.5mm} $P_\text{1S,tot}^\mathrm{NNLO^{\star}}$ \hspace{1.5mm} \\
 & 3.2725 & 10.018 &  2.913 & 5.16 & 21.36 & & 97.12 & 23.40 & 13.87 & 12.07 & 1.66 & 2.74 & 0.83 & 151.69 \\
\hline
\end{tabular}
\end{table}

To visualize the relative size of the real and virtual higher-order corrections, Table~\ref{tab:para_lo_nlo_nnlo_decay_widths} displays the partial annihilation decay widths of para-ditauonium grouped by LO, NLO, and NNLO contributions, where 
\begin{eqnarray}
\Gamma^\text{NLO}_{\gaga}\!&\!=\!&\!\Gamma^{(0)}_{\gaga}+\Delta\Gamma^\text{NLO}_\text{virt.exch}+\Delta\Gamma^\text{NLO}_\text{Coul},\;\mbox{ and }\; \Gamma^\text{NNLO}_{\gaga}=\Gamma^\text{NLO}_{\gaga}+  \Delta\Gamma^\text{NNLO}_\text{Breit}\,,
\end{eqnarray}
and where $\Gamma^\text{NLO}_\text{Dalitz}$ and $\Gamma^\text{NNLO}_\text{2Dalitz}$ sum up, respectively, all individual Dalitz and double-Dalitz decays.  From these results, a few quantitative facts can be highlighted: (i) altogether, the higher-order corrections augment the total $1^1$S$_0$ annihilation decay width by $+5.0\%$, (ii) the virtual NLO\,$+$\,NNLO corrections increase the dominant diphoton decay by $+0.8\%$, and (iii) the single- and double-Dalitz decays occur with a combined $\mathcal{B}=3.25\%$ rate (over the total width including single-$\tau$ weak decays, given by Eq.~(\ref{eq:para_tot_width}) below). Although the impact of the higher-order effects computed here may seem relatively small, the presence of final states with charged particles can facilitate the first experimental detection of para-ditauonium as discussed in Refs.~\cite{dEnterria:2022ysg,DdEHSS}.

\begin{table}[htpb!]
\tabcolsep=4.mm
\centering
\caption{Numerical values of the partial para-ditauonium decay widths grouped by individual LO, NLO, and NNLO contributions to its total width. The last column gives the total NNLO para-ditauonium annihilation width. \label{tab:para_lo_nlo_nnlo_decay_widths}}
\vspace{0.1cm}
\begin{tabular}{l|c|cc|cc|c}\hline
$\tata$ state &  $\Gamma^{(0)}_{\gaga}$ & $\Gamma^\text{NLO}_{\gaga}$ & $\Gamma^\text{NLO}_\text{Dalitz}$ & $\Gamma_{\gaga}^\text{NNLO}$ &  $\Gamma_\text{2Dalitz}^{\text{NNLO}}$ &  $\Gamma_{\gaga+\text{Dalitz}+\text{2Dalitz}}^{\text{NNLO}}$ \\ 
$1^1\mathrm{S}_0$ & 0.018384~eV & 0.018524~eV & 0.000772~eV & 0.018533~eV &  $5.42\cdot 10^{-6}$~eV & 0.01931~eV \\\hline
\end{tabular}
\end{table}

Table~\ref{tab:para_1_summary} summarizes all the properties of the $1^1\mathrm{S}_0$ state including all individual partial widths, up to the highest NLO\,$+$\,NNLO$^\star$ accuracy computed here, and associated $\mathcal{B}_{X} = \Gamma_X/\Gamma_\text{tot}$ branching fractions. The total ditauonium width is determined by adding all individual partial widths plus the effective width due to the weak tau decay, \ie\
\begin{equation}
\Gamma_\text{tot} = 
\Gamma^\text{NNLO}_{\gaga} + 
\Gamma^\text{NLO}_\text{Dalitz} +
\Gamma^\text{NNLO}_\text{2Dalitz} + 
\Gamma_{(2)\tau\to X},
\label{eq:para_tot_width}
\end{equation}

\begin{table}[htpb!]
\tabcolsep=3.5mm
\centering
\caption{Main properties (mass $m_X$, $J^{PC}$ quantum numbers, total width $\Gamma_\text{tot}$, lifetime, as well as partial decay widths $\Gamma_{X}$ and associated $\mathcal{B}_{X}$ branching fractions) of the lowest-energy para-ditauonium bound state computed in this work.\label{tab:para_1_summary}}
\vspace{0.1cm}
\begin{tabular}{lcccc|lcc}\toprule
$\tata$ state & $m_X$ (MeV) &  $J^{PC}$  &  $\Gamma_\text{tot}$ (eV)  &  Lifetime (fs)  & Decay mode  &  $\Gamma_{X}$ (eV)  &  $\mathcal{B}_{X}$  \\\midrule
$1^1\mathrm{S}_0$ & \;\;\texttt{$3553.696 \pm 0.240$}\;\; 
& $0^{-+}$ &  \texttt{$0.02384$} & 27.60
& $\gaga$ & \texttt{$0.018533$} & $77.72\%$ \\
 &  & & & & $\gamma\epem$  & \texttt{$4.28\cdot 10^{-4}$} & $1.79 \%$ \\
 &  & & & & $\gamma\mumu$  & \texttt{$1.24\cdot 10^{-4}$} & $0.52 \%$ \\
 &  & & & & $\gamma\qqbar$ & \texttt{$2.20\cdot 10^{-4}$} & $0.92 \%$ \\
 &  & & & & $\epem\epem$ & \texttt{$2.32\cdot 10^{-6}$} & $0.0094\%$\\
 &  & & & & $\epem\mumu$ & \texttt{$1.38\cdot 10^{-6}$} & $0.0058\%$\\
 &  & & & & $\epem\qqbar$ & \texttt{$1.20\cdot 10^{-6}$} & $0.0050\%$\\
 &  & & & & $\mumu\mumu$ & \texttt{$1.65\cdot 10^{-7}$} & $0.00069\%$\\
 &  & & & & $\mumu\qqbar$ & \texttt{$2.72\cdot 10^{-7}$} & $0.0011\%$\\
 &  & & & & $\qqbar\qqbarprime$ & \texttt{$8.23\cdot 10^{-8}$} & $0.00035\%$\\
 &  & & & & $(2)\tau\to X$ & \texttt{$0.004535$} & $19.02\%$ \\ 
\bottomrule
\end{tabular}
\end{table}

The last significant figures of all values listed in Table~\ref{tab:para_1_summary} have been rounded off to approximately match the associated theoretical accuracy of each width. Theoretical uncertainties due to missing higher-order corrections are very small for the total width and main diphoton decay channel, as they have been computed here including up to the most important NNLO corrections. The uncertainty is therefore at the NNLO level, $(\alpha/\pi)^2 \approx 10^{-4}$ accounting for $\mathcal{O}(10)$ coefficient prefactors (this is a realistic order-of-magnitude estimate, although partial cancellations between the complete set of NNLO virtual and real corrections, which have not been fully computed here, are not excluded). For the Dalitz $\gamma\ffbar$ para-ditauonium decays, since they are LO for this mode, their relative uncertainty is $\mathcal{O}(\alpha/\pi) \approx 10^{-2}$. The propagated parametric uncertainty due to the tau mass precision is around $7\cdot10^{-5}$ for all quantities, which linearly depend on $m_\tau$ through the LO widths. The uncertainty of the tau decay width due to the tau lifetime, Eq.~(\ref{eq:taudecaywidth}), is around $2\cdot 10^{-3}$, which propagates into $4\cdot10^{-4}$ relative uncertainty of the para-ditauonium total width. As one can see, theoretical uncertainties (intrinsic and parametric) are very small and very likely well beyond the reach of any potential experimental precision.

\subsection{Ortho-ditauonium decays}

\subsubsection{Real and virtual NLO and (partial) NNLO corrections}

\begin{figure}[htpb!]
\centering
\includegraphics[width=0.95\textwidth]{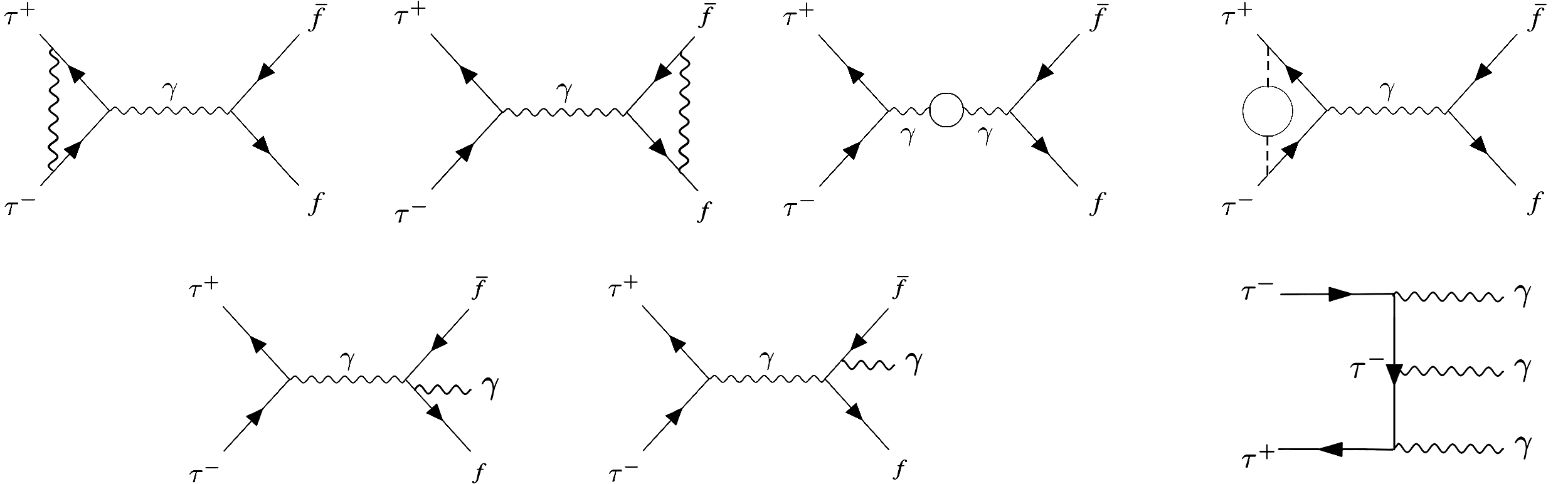}
\caption{Higher-order corrections to ortho-ditauonium $1^3{\text{S}}_1$ decays. The five leftmost diagrams show NLO virtual (top) and real (bottom) bremsstrahlung corrections to the $1^3{\text{S}}_1$ difermion decays. Top right: Wavefunction at the origin corrections to the $1^3{\text{S}}_1$ difermion decays. Bottom right: 3-photon decay.
\label{fig:NLO_ortho_diags}}
\end{figure}

Figure~\ref{fig:NLO_ortho_diags} shows the most important virtual and real higher-order QED corrections to the LO ortho-ditauonium decay width, $\Gamma(1 ^3\mathrm{S}_1)$. They include virtual photon exchanges (two top left panels), vacuum polarization loops of the annihilating virtual photon (top third panel), and modifications of order $\mathcal{O}(\alpha/\pi)$ of the ditauonium wavefunction at the origin (top right panel) affecting the difermion partial width, $\Gamma(1 ^3\mathrm{S}_1)\to\ffbar$. The bottom panels of the figure show real radiative corrections (left and center) of the difermion decays, and the 3-photon decay channel (right).

Let us deal first with the NLO virtual corrections to the $1 ^3{\text{S}}_1$ wavefunction at the origin (Fig.~\ref{fig:NLO_ortho_diags}, top right), which can be calculated through fermion loop insertions in the Coulomb's photon propagator, as done for the para-ditauonium case in Eq.~(\ref{eq:NLO_Coul}),
\begin{eqnarray}
\Delta\Gamma^\text{NLO}_{\text{Coul}}(1^3\mathrm{S}_1\to \ffbar)\!&\!=\!&\!\frac{\Delta|\varphi_{1\mathrm{S}}(0)|^2}{|\varphi_{1\mathrm{S}}(0)|^2}\Gamma^{(0)}(1 ^3\mathrm{S}_1\to \ffbar)=C_\text{Coul}^\text{NLO,1S}\, \left(\frac{\alpha}{\pi}\right)\, \Gamma^{(0)}(1 ^3\mathrm{S}_1\to \ffbar).
\label{eq:ortho_NLO_Coul}
\end{eqnarray}
Here, the $C_\text{Coul}^\text{NLO,1S} = 5.805$ coefficient has the same numerical value as computed before, Eq.~(\ref{eq:bound_green}).

We address next the ortho-ditauonium radiative corrections. The two $\gamma$-emission diagrams in Fig.~\ref{fig:NLO_ortho_diags} (bottom left and center) produce divergent infrared double logarithms 
that need to be canceled out against similar terms produced by the virtual correction diagrams shown in the three first top panels of the figure. 
The real and virtual radiative corrections can be calculated with the standard techniques by defining $x_{f^\prime}=m_{f^\prime}^2/m_\tau^2$, and ignoring power corrections of the form $\mathcal{O}(\alpha x_f)$, as follows:
\begin{eqnarray}
      \Delta\Gamma^\text{NLO}_{\text{rad.}}(1 ^3\mathrm{S}_1\to \ffbar(\gamma))
      \!&\!=\!&\!\left\{-\frac{13}{4}+\frac{1}{3}\sum_{f^\prime}{Q_{f^\prime}^2N_{c,f^\prime}\left[\left(2+x_{f^\prime}\right)\Re{\left(\sqrt{1-x_{f^\prime}}\ln{\left(\frac{\sqrt{1-x_{f^\prime}}+1}{\sqrt{1-x_{f^\prime}}-1}\right)}\right)}-\left(\frac{10}{3}+2x_{f^\prime}\right)\right]}\right\}\nonumber\\
      \!&\!\!&\! \times\, \left(\frac{\alpha}{\pi}\right)\, \Gamma^{(0)}(1 ^3\mathrm{S}_1\to \ffbar) \nonumber\\
      \!&\!=\!&\! \Bigg\{-\frac{13}{4}+\frac{1}{3}\sum_{\ell=e,\mu,\tau}{\left[\left(2+x_{\ell}\right)\Re{\left(\sqrt{1-x_{\ell}}\ln{\left(\frac{\sqrt{1-x_{\ell}}+1}{\sqrt{1-x_{\ell}}-1}\right)}\right)}-\left(\frac{10}{3}+2x_{\ell}\right)\right]}  + \!2\pi\frac{\Delta \alpha_\text{had}^{(3)}(m_{_{\tata}}^2)}{\alpha} \Bigg\}\nonumber\\
      \!&\!\!&\! \times\, \left(\frac{\alpha}{\pi}\right)\, \Gamma^{(0)}(1 ^3\mathrm{S}_1\to \ffbar) = C_\text{rad.}^\text{NLO}\,\left(\frac{\alpha}{\pi}\right)\,  \Gamma^{(0)}(1 ^3\mathrm{S}_1\to \ffbar),\,
      \mbox{ with }\, C_\text{rad.}^\text{NLO} = 15.860.
      \label{eq:NLOffX}
\end{eqnarray}
The first terms of this equation account for inclusive photon exchanges/emissions and leptonic vacuum polarization loops (the $\Re$ symbol indicates the real part of the expression in parenthesis), whereas the last term of the inner sum stands for the hadronic vacuum polarization contributions of the virtual photon at the ditauonium mass (\ie\ for the $N_f=3$ hadronic loop contributions of the third top diagram of Fig.~\ref{fig:NLO_ortho_diags}) quantified by the $\Delta \alpha_\text{had}^{(3)}(m_{_{\tata}}^2)=0.0077$ term 
(Table~\ref{tab:PDG}).


Equation~(\ref{eq:NLOffX}) determines the NLO contribution to the ortho-ditauonium width from photon emissions and exchanges inclusively. Here, next, we evaluate the size of the contributions from $1 ^3\mathrm{S}_1$ decays with photons explicitly tagged or measured in the final state, shown in the bottom (left and center) diagrams of Fig.~\ref{fig:NLO_ortho_diags}, where a photon energy cutoff needs to be introduced to avoid infrared divergences. We restrict ourselves to the case of the dilepton decays\footnote{Photon emission in ortho-ditauonium diquark decays are not considered here given that, first, they are suppressed by the square of the smaller charges of the quarks compared to the leptons, and are further complicated by the hadronization of the quarks that prevent a perturbative calculation of a hadronic final state including a potential quark-emitted photon.} with one final-state photon, $1 ^3\mathrm{S}_1\to \lele\gamma$. Collinear divergences can be regularised by keeping the outgoing lepton masses nonzero, while a threshold on the photon energy in the ditauonium rest frame is required to remove soft divergences, which we take as $E_{\gamma}>E^\text{thr}_{\gamma} \approx \mathcal{O}(m_\ell)$. By introducing two variables, $x_\ell=m_\ell^2/m_\tau^2$ and $\epsilon=E^\text{thr}_{\gamma}/m_\tau$, 
the partial width then reads,
\begin{eqnarray}
\hspace{-2mm}\Gamma^{(0)}(1 ^3\mathrm{S}_1\to \lele\gamma) =  \left[\frac{11}{4}\!-\!\frac{\pi^2}{3}\!+(2-4\ln{2})\ln{\epsilon}+(3/2+2\ln{\epsilon})\ln{x_\ell}-3\ln{2}+\mathcal{O}(\epsilon,x_\ell)\right]\!\left(\frac{\alpha}{\pi}\right)\! \Gamma^{(0)}(1^3\mathrm{S}_1\to \lele),
\end{eqnarray}
with the collinear $\ln{x_\ell}$ and soft $\ln{\epsilon}$ logarithms appearing, and where we have ignored  power corrections of order $\mathcal{O}(\epsilon,x_\ell)$. We have also computed the full expression with the $\epsilon$ and $x_\ell$ dependencies, which is a bit lengthy and we refrain ourselves from writing it here. The numerical values for $\epsilon=m_{\ell}/m_\tau$ are
\begin{eqnarray}
\Gamma^{(0)}(1 ^3\mathrm{S}_1\to \lele\gamma) &=& C^\text{NLO}_{\lele\gamma} \left(\frac{\alpha}{\pi}\right)\, \Gamma^{(0)}(1 ^3\mathrm{S}_1\to \lele),\; \mbox{ with }\;
C^\text{NLO}_{\lele\gamma} = \left\{\begin{array}{lcl} 245.13 & {\rm for} & \ell=e,\\
23.76 & {\rm for} & \ell=\mu.\\\end{array}\right.\label{eq:widthllgamma}
\end{eqnarray}
To estimate the size of the purely virtual NLO corrections of ortho-ditauonium, we can combine Eqs.~(\ref{eq:NLOffX}) and~(\ref{eq:widthllgamma}) and determine the NLO partial width for $1^3\mathrm{S}_1\to \lele$ \textit{without} real photon emission (with an energy above $m_\ell$), as follows,
\begin{eqnarray}
\Gamma^{\rm NLO}(1 ^3\mathrm{S}_1\to \lele)\!&\!=\!&\!\left[\Gamma^{(0)}(1 ^3\mathrm{S}_1\to \lele)+\Delta\Gamma^\text{NLO}_{\text{rad.}}(1 ^3\mathrm{S}_1\to \lele(\gamma))+\Delta\Gamma^\text{NLO}_{\text{Coul}}(1 ^3\mathrm{S}_1\to \lele)-\Gamma^{(0)}(1 ^3\mathrm{S}_1\to \lele\gamma)\right]\nonumber\\
      \!&\!=\!&\!
     \left[1+\left(\frac{\alpha}{\pi}\right)C^\text{NLO}_{\text{virt.exch},\ell}\right]\,\Gamma^{(0)}(1 ^3\mathrm{S}_1\to \lele)
     ,\; \mbox{ with }\; C^\text{NLO}_{\text{virt.exch},\ell} = \left\{\begin{array}{lcl}-223.47 & {\rm for} & \ell=e,\\
      -2.092 & {\rm for} & \ell=\mu.\\\end{array}\right.
\end{eqnarray}
The leading NNLO virtual contribution from the wavefunction correction from the Breit Hamiltonian~\cite{Berestetskii:1982qgu} is
\begin{eqnarray}
\Delta \Gamma^\text{NNLO}_\text{Breit}(n^3\mathrm{S}_1\to \ffbar)\!&\!=\!&\!C^\text{NNLO}_\text{Breit}\left(\frac{\alpha}{\pi}\right)^2\Gamma^{(0)}(n^3\mathrm{S}_1\to \ffbar),\; \mbox{ with }\; C^\text{NNLO}_\text{Breit} = -\frac{\pi^2}{3}\ln{\left(1/\alpha\right)} = -16.187.\label{eq:BreitNNLOortho}
\end{eqnarray}
This (small) NNLO virtual correction is negative and will slightly \textit{decrease} the partial $\Gamma_{\ffbar(\gamma)}$ decay widths although, as we see next, there are extra real NLO\,$+$\,NNLO contributions (3-photons and 4-fermions final states) that contribute positively to the total ortho-ditauonium decay rate.

Figure~\ref{fig:NLO_ortho_diags} (bottom right) shows the 3-photon decay channel of ortho-ditauonium, which is suppressed by an extra $\alpha$ factor compared to the diphoton para-ditauonium decay given by Eq.~(\ref{eq:gaga_tata_width}). At its lowest order, this partial decay width is
\begin{equation}
\Gamma^{(0)}(n {}^3\mathrm{S}_1 \to\gaga\gamma) = \frac{2(\pi^2 - 9)\,\alpha^6 m_\tau}{9\pi\,n^3} = C_{3\gamma}\left(\frac{\alpha}{\pi}\right)\Gamma^{(0)}(n{}^3\mathrm{S}_1),\; \mbox{ with }\; C_{3\gamma} = \frac{4(\pi^2-9)}{3\cdot 4.2} = 0.2761.\\
\label{eq:3gammaLO}
\end{equation}
The (infrared-finite) NLO virtual contribution of this channel can be calculated by the standard perturbative techniques, and amounts to
\begin{eqnarray}
\Delta\Gamma^\text{NLO}_{\text{virt.exch}}(n ^3\mathrm{S}_1\to\gaga\gamma)\!&\!=\!&\! -13.44\,\left(\frac{\alpha}{\pi}\right)\Gamma^\text{(0)}(n ^3\mathrm{S}_1\to\gaga\gamma)\nonumber\\
\!&\!=\!&\!C^\text{NLO}_{{\rm virt.exch},3\gamma}\left(\frac{\alpha}{\pi}\right)^2\Gamma^{(0)}(n{}^3\mathrm{S}_1),\label{eq:NLOaaa_Vert} \; \mbox{ with }\; C^\text{NLO}_{{\rm virt.exch},3\gamma} = -13.44\cdot C_{3\gamma} = -3.71.
\end{eqnarray}
The NLO correction from the wavefunction at the origin for the 3-$\gamma$ decay is
\begin{eqnarray}
\Delta\Gamma^\text{NLO}_{\text{Coul}}(n^3\mathrm{S}_1\to\gaga\gamma)\!&\!=\!&\!C_\text{Coul}^{\text{NLO},n\mathrm{S}}\,\left(\frac{\alpha}{\pi}\right)\Gamma^\text{(0)}(n ^3\mathrm{S}_1\to\gaga\gamma)\nonumber\\
\!&\!=\!&\!C^{\text{NLO},n\mathrm{S}}_{\text{Coul},3\gamma}\left(\frac{\alpha}{\pi}\right)^2\Gamma^{(0)}(n{}^3\mathrm{S}_1)\,, \; \mbox{ with }\; C^{\text{NLO},n\mathrm{S}}_{\text{Coul},3\gamma}=C_\text{Coul}^{\text{NLO},n\mathrm{S}}\cdot C_{3\gamma} = 1.6026\; \mbox{ for }\; n=1.
\label{eq:NLOaaa_Coul}
\end{eqnarray}
One can see that the net effect of the NLO corrections to the 3-photon decay, combining Eqs.~(\ref{eq:NLOaaa_Vert}) and (\ref{eq:NLOaaa_Coul}), leads to a small \textit{decrease} (by about $-1.8\%$) of this rate.

\begin{figure}[htpb!]
\centering
\includegraphics[width=0.7\textwidth]{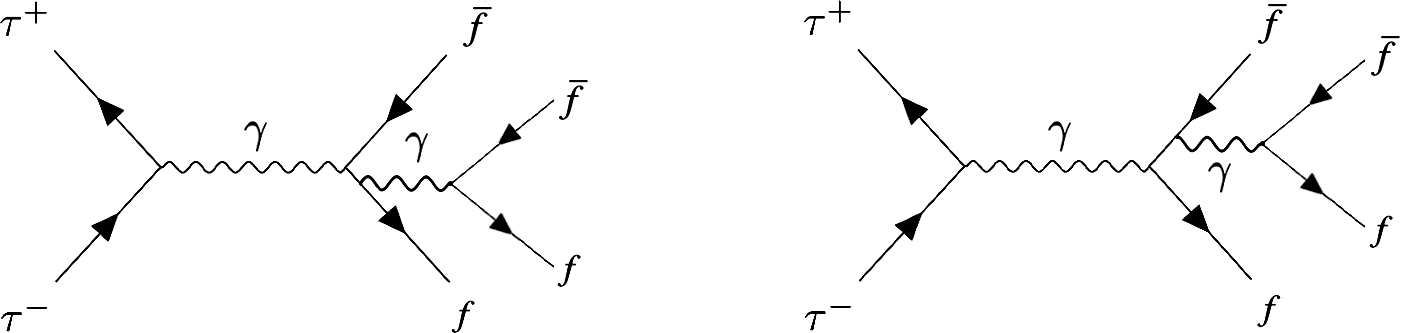}
\caption{Representative Feynman diagrams of (real NNLO) ortho-ditauonium decays into four fermions.
\label{fig:ortho_4fermions_diags}}
\end{figure}

Finally, the ortho-ditauonium can also have 4-fermion decay channels contributing at real NNLO accuracy to the total width (Fig.~\ref{fig:ortho_4fermions_diags}). The partial widths can be again cast into the following generic form
\begin{equation}\label{eq:ortho4f}
\Gamma^\text{NNLO}_{4f}(n{}^3\mathrm{S}_1\to\ffbar\ffbarprime) = C^\text{NNLO}_{4f}(\ffbar\ffbarprime) \left(\frac{\alpha}{\pi}\right)^2\Gamma^{(0)}(n{}^3\mathrm{S}_1)\,,
\end{equation}
with the numerical values of the $C^\text{NNLO}_{4f}(\ffbar\ffbarprime)$ coefficients, derived with the help of {\sc\small HELAC-Onia}~\cite{Shao:2012iz,Shao:2015vga}, given by
\begin{eqnarray}
      C^\text{NNLO}_{4f}(\epem\epem)\!&\!=\!&\!40.0,\quad C^\text{NNLO}_{4f}(\epem\mumu)\!=\!30.3,\quad 
      C^\text{NNLO}_{4f}(\epem\qqbar)\!=\!13.3,\nonumber\\
      C^\text{NNLO}_{4f}(\mumu\mumu)\!&\!=\!&\!0.886,\;\; C^\text{NNLO}_{4f}(\mumu\qqbar)\!=\!0.530,\quad\; C^\text{NNLO}_{4f}(\qqbar\qqbarprime)\!=\!0.070.
      \label{eq:4f_results}
\end{eqnarray}

\subsubsection{Rare weak decays}
\label{sec:nunu}

\begin{figure}[htpb!]
\centering
\includegraphics[width=0.65\textwidth]{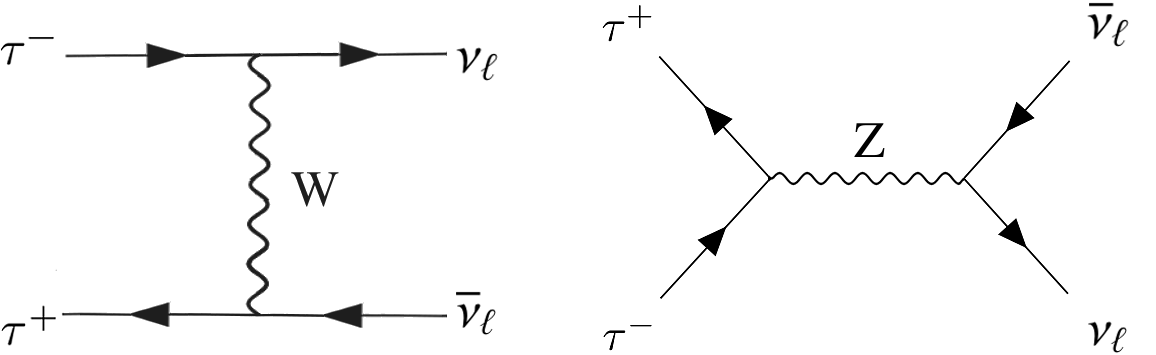}
\caption{LO diagrams of ortho-ditauonium decays into neutrinos.
\label{fig:ortho_nunu_diags}}
\end{figure}

Of the two spin states of ditauonium, only\footnote{This holds assuming that neutrinos are massless, and without emission of photons in the potential para-ditauonium decay process.} the ortho state can decay weakly into a pair of neutrinos. The two amplitudes that contribute to the decay $\tataOne \to \nu_{\ell}\overline{\nu}_{\ell}$, are W exchange in the $t$ channel and Z annihilation in the $s$-channel (Fig.~\ref{fig:ortho_nunu_diags}). The Z diagram is a tiny correction to the virtual photon annihilation (Fig.~\ref{fig:LO_decays_diags}, right), and there is a destructive interference between the W and Z exchange amplitudes. The corresponding partial widths were found negligible for positronium and dimuonium ---for positronium they are $\mathcal{O}(10^{-18})$ and $\mathcal{O}(10^{-21})$ for decays into like- and unlike-flavour between the $\epem$ and neutrinos respectively~\cite{Czarnecki:1999mt}---, but they are comparatively enhanced by powers of the tau to electron (or muon) masses for ditauonium, and it is worth to estimate their importance here.

For ortho-ditauonium, the neutrino-pair decay widths read
\begin{eqnarray}
\Gamma^{(0)}(n^3\mathrm{S}_1\to \nu_\tau\bar{\nu}_\tau)\!&\!=\!&\!\frac{\alpha^5m_\tau}{6n^3}\frac{m_\tau^4\left[m_\mathrm{W}^2\left(m_\mathrm{W}^2+m_\tau^2\right)\left(1-4s_w^2\right)-\left(2m_\mathrm{W}^2+m_\tau^2\right)\left(m_\mathrm{Z}^2-4m_\tau^2\right)c_w^2\right]^2}{8s_w^4c_w^4m_\mathrm{W}^4\left(m_\mathrm{W}^2+m_\tau^2\right)^2\left(m_\mathrm{Z}^2-4m_\tau^2\right)^2}\nonumber\\
\!&\!=\!&\!2.147\cdot 10^{-6}\cdot \Gamma^{(0)}(n^3\mathrm{S}_1\to\epem),\nonumber\\
\Gamma^{(0)}(n^3\mathrm{S}_1\to \nu_e\bar{\nu}_e,\nu_\mu\bar{\nu}_\mu)\!&\!=\!&\!\frac{\alpha^5m_\tau}{6n^3}\frac{m_\tau^4\left(1-4s_w^2\right)^2}{8s_w^4c_w^4\left(m_\mathrm{Z}^2-4m_\tau^2\right)^2}=7.015\cdot 10^{-9}\cdot\Gamma^{(0)}(n^3\mathrm{S}_1\to\epem),\nonumber
\end{eqnarray}
where $m_\mathrm{W,Z}$ are the W and Z boson masses, and $s_w$ and $c_w$ are the sine and cosine of the Weinberg angle. Both channels turn out to be also very rare for ditauonium due to the strong suppression driven by the $m_\tau^4/m_\mathrm{W,Z}^4$ factor.

\subsubsection{Combined higher-order corrections}

The total annihilation decay width for $n=1$ ortho-ditauonium can be now obtained from all previous results and written in the compact form
\begin{equation}\label{eq:ortho_width_corrs}
\Gamma_\text{tot}(1^3\mathrm{S}_1)=\left(1+O_\text{1S,tot}^\text{NLO}\left(\frac{\alpha}{\pi}\right)+O_\text{1S,tot}^\mathrm{NNLO^{\star}}\left(\frac{\alpha}{\pi}\right)^2\right)\Gamma^{(0)}(1 ^3\mathrm{S}_1),
\end{equation}
where $\Gamma^{(0)}(1 ^3\mathrm{S}_1)$ is given in Eq.~(\ref{eq:LOwidthortho}). The coefficient $O_\text{1S,tot}^\text{NLO}$ sums up all numerical NLO coefficients derived in Eqs.~(\ref{eq:ortho_NLO_Coul}),~(\ref{eq:NLOffX}), and~(\ref{eq:3gammaLO}), namely
\begin{eqnarray}
O_\text{1S,tot}^\text{NLO}= C_\text{Coul}^\text{NLO,1S} + C_\text{rad.}^\text{NLO} + C_{3\gamma} = 5.805 + 15.860 + 0.276 = 21.94\,,
\label{eq:ortho1tot}
\end{eqnarray}
whereas $O_\text{1S,tot}^\mathrm{NNLO^{\star}}$ sums up the partial NNLO corrections given by Eqs.~(\ref{eq:BreitNNLOortho}),~(\ref{eq:NLOaaa_Vert}), (\ref{eq:NLOaaa_Coul}), (\ref{eq:ortho4f}), and (\ref{eq:4f_results}), as follows
\begin{equation}\label{eq:ortho1tot1}
O_\text{1S,tot}^\mathrm{NNLO^{\star}}=C_\text{Breit}^\text{NNLO}+C^\text{NLO}_{\text{virt.exch},3\gamma}+C^{\text{NLO},1\mathrm{S}}_{\text{Coul},3\gamma}+\sum_{f,f^\prime}{C_{4f}^\text{NNLO}(\ffbar\ffbarprime)}=-16.19-3.71+1.60+85.1 = 66.8\,.
\end{equation}
The corresponding values of all individual coefficients are listed in Table~\ref{tab:orthoditauonium_corrs}. Compared to the para-ditauonium case, one can see that the size of the NLO corrections are numerically very similar ---the $(\alpha/\pi)$ prefactor coefficients are $P_\text{1S,tot}^\text{NLO} = 21.36$ from Eq.~(\ref{eq:p1tot}) vs.\ $O_\text{1S,tot}^\text{NLO}= 21.94$ from Eq.~(\ref{eq:ortho1tot}), respectively---, whereas the (small) NNLO corrections are larger for the para- than for the ortho-state as one can see comparing Eqs.~(\ref{eq:para_nnlo_tot}) and (\ref{eq:ortho1tot1}). This implies that the total NLO\,$+$\,NNLO$^\star$ corrections increase by almost the same amount, about $5.0\%$, the LO annihilation rates of both states.

\begin{table}[htbp!]
\tabcolsep=4.5mm
\centering
\caption{Coefficients for the sum of all NLO and NNLO corrections of each partial ortho-ditauonium decay channel, $C^\text{NLO,NNLO}_\text{part}$, and final $O^\text{NLO}_\text{1S,tot}$ and $O_\text{1S,tot}^\mathrm{NNLO^{\star}}$ correction terms as defined in Eqs.~(\ref{eq:ortho1tot}) and~(\ref{eq:ortho1tot1}) for its total annihilation decay width.  The numerical values for the exclusive $\epem\gamma$ and $\mumu\gamma$ modes are given for different threshold photon energies, $E_\gamma^\text{thr} > m_{e,\mu}$ respectively (see text for details).
\label{tab:orthoditauonium_corrs}}
\vspace{0.1cm}
\begin{tabular}{cccccccc}\hline 
$C^\text{NLO}_\text{part}:$ & $\epem$  & $\mumu$ & $\qqbar(\gamma)$ & $\gaga\gamma$ & $\epem\gamma$ & $\mumu\gamma$ & $O_\text{1S,tot}^\text{NLO}$ \\
& $-53.21$ & $-0.498$ & 11.348 & 0.276 & 58.364 & 5.657 & 21.94 \\\hline
$C^\text{NNLO}_\text{part}:$ &  $\epem$ & $\mumu$ & $\qqbar$ & $\gamma\gamma\gamma$ & $4f$ & & $O_\text{1S,tot}^\mathrm{NNLO^{\star}}$  \\
 & $-3.854$ &  $-3.854$ & $-8.479$ & $-2.107$ & $85.1$ & & 66.8 \\
\hline
\end{tabular}
\end{table}

\begin{table}[htpb!]
\tabcolsep=2.mm
\centering
\caption{Numerical values of the partial ortho-ditauonium decay widths (in~eV) grouped by individual LO, NLO, and NNLO contributions to its total width. The last column gives the total NNLO$^\star$ ortho-ditauonium annihilation width. For the $\lele(\gamma)$ partial widths, dielectron and dimuon channels are added up. \label{tab:ortho_lo_nlo_nnlo_decay_widths}}
\vspace{0.1cm}
\begin{tabular}{l|cc|ccc|cccc|c}\hline
$\tata$ state & $\Gamma^{(0)}_{\lele}$ & $\Gamma^{(0)}_{\qqbar}$ & $\Gamma^\text{NLO}_{\lele(\gamma)}$ & $\Gamma^\text{NLO}_{\qqbar(\gamma)}$ & $\Gamma^{(0)}_{3\gamma}$ & $\Gamma_{\lele(\gamma)}^\text{NNLO}$ &  $\Gamma_{\qqbar(\gamma)}^{\text{NNLO}}$ & $\Gamma_{3\gamma}^\text{NLO}$ & $\Gamma_{4f}^\text{NNLO}$ & $\Gamma_{\ffbar(\gamma)+3\gamma+4f}^{\text{NNLO}}$ \\ \hline
$1^1\mathrm{S}_0$ & 0.01226 & 0.0135 & 0.0129 & 0.0142 & $1.65\cdot 10^{-5}$ & 0.0129 & 0.0142 & $1.62\cdot 10^{-5}$ & $1.18\cdot 10^{-5}$ & 0.02706 \\\hline
\end{tabular}
\end{table}

Table~\ref{tab:ortho_lo_nlo_nnlo_decay_widths} displays the partial annihilation decay widths of ortho-ditauonium grouped at LO, NLO, and NNLO accuracies to visualize the relative size of the real and virtual higher-order corrections, where we define
\begin{eqnarray}
\Gamma^\text{NLO}_{\lele(\gamma)}\!&\!=\!&\!\!\sum_{\ell=e,\mu}{\left[\Gamma^{(0)}(1 ^3\mathrm{S}_1\to\lele)+\Delta \Gamma^\text{NLO}_\text{rad.}(1 ^3\mathrm{S}_1\to\lele(\gamma))+\Delta \Gamma^\text{NLO}_\text{Coul}(1 ^3\mathrm{S}_1\to\lele)\right]},\nonumber\\
\Gamma^\text{NNLO}_{\lele(\gamma)}\!&\!=\!&\!\Gamma^\text{NLO}_{\lele(\gamma)}+\sum_{\ell=e,\mu}{\Delta \Gamma^\text{NNLO}_\text{Breit}(1 ^3\mathrm{S}_1\to\lele)},\nonumber\\
\Gamma^\text{NLO}_{\qqbar(\gamma)}\!&\!=\!&\!\sum_{q=u,d,s}{\left[\Gamma^{(0)}(1 ^3\mathrm{S}_1\to\qqbar)+\Delta \Gamma^\text{NLO}_\text{rad.}(1 ^3\mathrm{S}_1\to\qqbar(\gamma))+\Delta \Gamma^\text{NLO}_\text{Coul}(1 ^3\mathrm{S}_1\to\qqbar)\right]},\\
\Gamma^\text{NNLO}_{\qqbar(\gamma)}\!&\!=\!&\!\Gamma^\text{NLO}_{\qqbar(\gamma)}+\sum_{q=u,d,s}{\Delta \Gamma^\text{NNLO}_\text{Breit}(1 ^3\mathrm{S}_1\to\qqbar)},\nonumber\\
\Gamma^\text{NLO}_{3\gamma}\!&\!=\!&\!\Gamma^{(0)}(1 ^3\mathrm{S}_1\to\gaga\gamma) + \Delta\Gamma^\text{NLO}_\text{virt.exch}(1 ^3\mathrm{S}_1\to\gaga\gamma)+ \Delta\Gamma^\text{NLO}_\text{Coul}(1 ^3\mathrm{S}_1\to\gaga\gamma),\nonumber\\
\Gamma^\text{NNLO}_{4f}\!&\!=\!&\!\sum_{f,f^\prime}{\Gamma^{(0)}(1 ^3\mathrm{S}_1\to\ffbar\ffbarprime)}\,.\nonumber
\end{eqnarray}

Table~\ref{tab:ortho_1_summary} lists all the properties of the $1^3\mathrm{S}_1$ state computed here. The total ortho-ditauonium width listed is determined by adding all individual partial widths plus the effective width due to the constituent tau weak decays, \ie\
\begin{equation}
\Gamma_\text{tot} = \Gamma^\text{NNLO}_{\lele(\gamma)}+\Gamma^\text{NNLO}_{\qqbar(\gamma)}+\Gamma^\text{NLO}_{3\gamma}+\Gamma^\text{NNLO}_{4f}+\Gamma_{(2)\tau\to X}\,,
\end{equation}

Our main findings about ortho-ditauonium decay rates can be summarized as follows: (i) the radiative real and virtual NLO corrections increase the difermion decays by $+5.0\%$, (ii) the virtual NNLO corrections are tiny and negative but ``compensated'' in the total width by positive real NNLO corrections from new 4-fermion channels that open up at this level of accuracy (with combined branching fractions of $\mathcal{B}\approx 0.04\%$), and (iii) the decays into a pair of neutrinos have $\mathcal{O}(10^{-7}\mbox{--}10^{-9})$ rates. The ortho-ditauonium branching fractions are thus dominated by decays into a pair of light diquarks (with or without $\gamma$ emission), $\mathcal{B}_{\qqbar(\gamma)} = 44.82\%$, with the actual hadronic final states mostly consisting of a few charged and neutral pions and/or, to a less extent, kaons. The combined dilepton final states with or without photon emission, $1^3\text{S}_1 \to\epem(\gamma), \mumu(\gamma)$, have a branching fraction of $\mathcal{B}_{\lele(\gamma)} = 2\times 20.37\% = 40.74\%$. The presence in the ortho-ditauonium decays of final states with different charged particles can facilitate the measurement of this exotic atom, given that the experimental momentum and vertex resolutions are better for them than for photons~\cite{dEnterria:2022ysg,DdEHSS}.

\begin{table}[htpb!]
\tabcolsep=3.mm
\centering
\caption{Main properties (mass $m_X$, $J^{PC}$ quantum numbers, total width $\Gamma_\text{tot}$, lifetime, as well as partial decay widths $\Gamma_{X}$ and associated $\mathcal{B}_{X}$ branching fractions) of the lowest-energy ortho-ditauonium bound state computed in this work. The two inclusive decay modes $\lele(\gamma)$ have been also broken-down into the exclusive modes $\lele$ and $\lele\gamma$ (where the latter require threshold photon energies $E_\gamma^\text{thr} > m_{e,\mu}$ for the dielectron and dimuon channels, respectively, see text for details). \label{tab:ortho_1_summary}}
\vspace{0.1cm}
\begin{tabular}{lcccc|lcc}\toprule
$\tata$ state & $m_X$ (MeV) &  $J^{PC}$  &  $\Gamma_\text{tot}$ (eV)  &  Lifetime (fs)  & Decay mode  &  $\Gamma_{X}$ (eV)  &  $\mathcal{B}_{X}$  \\\midrule
$1^3\mathrm{S}_1$ & \;\;\texttt{$3553.696 \pm 0.240$}\;\; & $1^{--}$ & \texttt{$0.03159$} & 20.83 & 
$\epem(\gamma)$ & \texttt{$0.006436$} & $20.37\%$ \\
 & & & & & $\;\circ\;\;$ $\epem$ & \;\;\;\;\;\;\texttt{$2.95\cdot 10^{-3}$} & \;\;\;\;\;\;$9.33\%$ \\
 &  & & & & $\;\circ\;\;$ $\epem\gamma$ & \;\;\;\;\;\;\texttt{$3.49\cdot 10^{-3}$} & \;\;\;\;\;\;$11.04\%$ \\
 &  & & & & $\mumu(\gamma)$ & \texttt{$0.006436$} & $20.37\%$ \\
  &  & & & & $\;\circ\;\;$ $\mumu$ & \;\;\;\;\;\;\texttt{$6.10\cdot 10^{-3}$} & \;\;\;\;\;\;$19.30\%$\\
 &  & & & & $\;\circ\;\;$ $\mumu\gamma$ & \;\;\;\;\;\;\texttt{$3.38\cdot 10^{-4}$} & \;\;\;\;\;\;$1.07\%$ \\
 &  & & & & $\qqbar(\gamma)$ & \texttt{$0.01416$} & $44.82\%$ \\
  &  & & & & $\gaga\gamma$ & \texttt{$1.62\cdot 10^{-5}$} & $0.051\%$ \\
%
 &  & & & & $\epem\epem$ & \texttt{$5.55\cdot 10^{-6}$} & $0.0176\%$ \\ 
 &  & & & & $\epem\mumu$ & \texttt{$4.21\cdot 10^{-6}$} & $0.0133\%$ \\ 
 &  & & & & $\epem\qqbar$ & \texttt{$1.85\cdot 10^{-6}$} & $0.0058\%$ \\
&  & & & & $\mumu\mumu$ & \texttt{$1.23\cdot 10^{-7}$} & $\mathcal{O}(10^{-6})$ \\ 
&  & & & & $\mumu\qqbar$ & \texttt{$7.36\cdot 10^{-8}$} & $\mathcal{O}(10^{-6})$ \\ 
&  & & & & $\qqbar\qqbarprime$ & \texttt{$9.73\cdot 10^{-9}$} & $\mathcal{O}(10^{-7})$ \\ 
 &  & & & & $\nu_\tau\bar{\nu}_\tau$ & \texttt{$1.32\cdot 10^{-8}$} & $\mathcal{O}(10^{-7})$ \\ 
 &  & & & & $\nu_{e}\bar{\nu}_{e}$ & \texttt{$4.30\cdot 10^{-11}$} & $\mathcal{O}(10^{-9})$ \\
  &  & & & & $\nu_{\mu}\bar{\nu}_{\mu}$ & \texttt{$4.30\cdot 10^{-11}$} & $\mathcal{O}(10^{-9})$ \\
 &  & & & & $(2)\tau\to X$ & \texttt{$0.004535$} & $14.35\%$ \\ 
\bottomrule
\end{tabular}
\end{table}

As discussed for the para-ditauonium case, theoretical uncertainties due to missing higher-order corrections for the total width and main difermion decay channels are very small, since we have computed them including up to the most important NNLO corrections. The uncertainty is therefore at the NNLO level, $\alpha^2 \approx 10^{-4}$ (this is a realistic order-of-magnitude
estimate, although partial cancellations between the complete set of NNLO virtual and real corrections, which have not been fully computed here, are not excluded). For the decay to hadrons, the perturbative uncertainty is also NNLO (\ie\ $10^{-4}$), but the uncertainty of $\Delta\alpha_\text{had}^{(3)}(m_{_{\tata}}^2)$ and $R_\text{had}(m_\tau^2)$ can propagate into a larger/dominant value. Since we have provided the analytic result, Eq.~(\ref{eq:LOwidthorthohad}), the numbers can be easily updated whenever the experimental value of $R_\text{had}(m_\tau^2)$ is refined. In all numerical evaluations, which linearly depend on $m_\tau$ via the LO widths, the propagated parametric uncertainty due to the tau mass precision is around $7\cdot10^{-5}$. The uncertainty in the tau decay width propagates into a $3\cdot10^{-4}$ relative uncertainty of the ortho-ditauonium total width. Therefore, theoretical uncertainties (both of intrinsic and parametric nature) are very small, and very likely well beyond the precision of any actual experimental measurement.

\section{Summary}
\label{sec:summ}

We have presented the first study of the spectroscopic structure of the purely leptonic system consisting of two $\tau$ leptons, bound by their mutual QED interaction, known as ditauonium. First, the basic zeroth-order expressions for its energy levels and dominant decay widths have been presented. The ground state (1S) has a binding energy of $-23.655$~keV, leading to a ditauonium mass of $m_{_{\tata}} = 3553.696 \pm 0.240$~MeV (where the uncertainty is dominated by the current tau lepton mass precision). 
Ditauonium decays dominantly through annihilations into pairs of photons and of lighter charged fermions for para- (1$^{1}$S$_{0}$) and ortho- (1$^{3}$S$_{1}$) states, respectively. Secondly, QED corrections at NLO and (partially) NNLO accuracy have been calculated for the energy levels and for the rates of all decay modes kinematically accessible at each level of accuracy. For the ground state, we find that the Lamb shift decreases its binding energy by about 0.115~keV, whereas the hyperfine splitting separates the para- and ortho-states by $\mathcal{O}(3$~eV).\\

A detailed study of all partial decay widths of para- and ortho-ditauonium has been carried out. Including all NLO and the most important NNLO corrections, the annihilation decay widths of $1^1\text{S}_0$ and $3^1\text{S}_1$ ditauonium states increase both by $+5.0\%$ compared to the LO results. The total decay widths of the para and ortho ground states (adding in both cases an effective width of $\Gamma_{(2)\tau\to X} = 0.004535$~eV from the weak decay of any of its constituent leptons) amount to $\Gamma^{\text{NNLO}^\star}_\text{tot} = 0.02384$~eV and $0.03159$~eV, corresponding to lifetimes of $\tauup = 27.60$~fs and 20.83~fs, respectively. In the para-ditauonium case, its leading diphoton decay mode receives $+0.8\%$ contributions from virtual NLO\,$+$\,NNLO corrections, for a total final branching fraction of $\mathcal{B}_{\gaga} = 77.72\%$. The real NLO corrections for this decay correspond to ``Dalitz'' final states, $1^1\text{S}_0 \to\gamma\epem, \gamma\mumu, \gamma\qqbar$ with branching ratios of $\mathcal{B}_{\gamma\ffbar} = 1.79\%, 0.52\%, 0.92\%$, respectively, whereas NNLO ``double Dalitz'' decays have a tiny $\mathcal{B}_{\ffbar\ffbarprime} = 0.02\%$ combined rate. The presence of final states with charged leptons, although at a few percent rate, can facilitate the experimental observation of para-ditauonium.
The calculated ortho-ditauonium branching fractions in dilepton final states with or without photon emission, $1^3\text{S}_1 \to\epem(\gamma), \mumu(\gamma)$, are $\mathcal{B}_{\lele(\gamma)} = 20.37\%$ each, about half the rate expected for similar decays with light diquarks, $\mathcal{B}_{\qqbar(\gamma)} = 44.82\%$. The existence of different charged-particle decay modes in the ortho-ditauonium decays can also facilitate the measurement of this exotic atom by exploiting the better experimental momentum and secondary vertex resolutions for $e^\pm,\mu^\pm,\pi^\pm,$ and K$^\pm$ compared to photons. Ortho-ditauonium decays into three photons or four fermions (combined) have tiny branching ratios of $\mathcal{B}_{3\gamma,\ffbar\ffbarprime} \approx 0.05\%$ each, and decays into neutrinos are at the $\mathcal{O}(10^{-7})$ level.\\

Ditauonium is the heaviest and most compact purely leptonic ``atomic'' system and remains experimentally unobserved to date. The results presented here are of usefulness to carry out potential experimental measurements of its production and study its properties at current and future colliders via multiple different final states whose decay rates have been quantified here for the first time. Concrete feasibility cases, beyond those described in~~\cite{dEnterria:2022ysg}, will be presented in an upcoming work~\cite{DdEHSS}. Ditauonium studies can provide novel tests of bound-state QED that are sensitive to physics beyond the standard model at higher energies than those of its lighter siblings positronium and dimuonium.\\


\paragraph*{Acknowledgments.---} Discussions with Bogdan Malaescu on the hadronic running of $\alpha_\text{QED}$ are gratefully acknowledged. Support from the European Union’s Horizon 2020 research and innovation program (grant agreement No.824093, STRONG-2020, EU Virtual Access ``NLOAccess''), the French ANR (grant ANR-20-CE31-0015, ``PrecisOnium''), and the CNRS IEA (grant No.205210, ``GlueGraph"), are acknowledged.


\bibliographystyle{myutphys}
\bibliography{reference}


\end{document}